\documentclass[11pt,twoside]{article}
\usepackage[utf8]{inputenc}

\usepackage[acronym,toc]{glossaries}
\makenoidxglossaries
\newglossaryentry{latex}
{
    name=latex,
    description={Is a mark up language specially suited 
    for scientific documents}
}

\newglossaryentry{maths}
{
    name=mathematics,
    description={Mathematics is what mathematicians do}
}

\newacronym{gcd}{GCD}{Greatest Common Divisor}
\newacronym{lcm}{LCM}{Least Common Multiple}
\newacronym{READS}{READS}{Accelerator Real-time Edge AI for Distributed Systems}
\newacronym{FNAL}{FNAL}{Fermi National Accelerator Laboratory}
\newacronym{DOE}{DOE}{Department of Energy}
\newacronym{AD}{AD}{Accelerator Division}
\newacronym{HEP}{HEP}{High Energy Physics}
\newacronym{LHC}{LHC}{Large Hadron Collider}
\newacronym{CMS}{CMS}{Compact Muon Solenoid}
\newacronym{DUNE}{DUNE}{Deep Underground Neutrino Experiment}
\newacronym{PIP-II}{PIP-II}{Proton Improvement Plan-II}
\newacronym{NU}{NU}{Northwestern University}
\newacronym{AI}{AI}{Artificial Intelligence}
\newacronym{ML}{ML}{Machine Learning}
\newacronym{NuMI}{NuMI}{Neutrinos at the Main Injector}
\newacronym{ACNET}{ACNET}{Accelerator Control Network}
\newacronym{TCLK}{TCLK}{Tevatron Clock}
\newacronym{MDAT}{MDAT}{Machine Data}
\newacronym{MCR}{MCR}{Main Control Room}
\newacronym{Mu2e}{Mu2e}{Muon to Electron Conversion Experiment}
\newacronym{DR}{DR}{Delivery Ring}
\newacronym{SDF}{SDF}{Spill Duty Factor}
\newacronym{RFKO}{RFKO}{Radio Frequency Knock Out}
\newacronym{PID}{PID}{Proportional-Integral-Derivative feedback loop}
\newacronym{POT}{POT}{Protons on Target}
\newacronym{RR}{RR}{Recycler Ring}
\newacronym{MI}{MI}{Main Injector}
\newacronym{BLM}{BLM}{Beam Loss Monitor}
\newacronym{FPGA}{FPGA}{Field Programmable Gate Array}
\newacronym{FM}{FM}{Frequency Modulated}
\newacronym{RF}{RF}{Radio Frequency}
\newacronym{ESS}{ESS}{Electrostatic Septum}
\newacronym{SRS}{SRS}{Spill Regulation System}
\newacronym{GPU}{GPU}{Graphics Processing Unit}
\newacronym{RTL}{RTL}{Register Transfer Level}
\newacronym{HLS}{HLS}{High Level Synthesis}
\newacronym{HDL}{HDL}{Hardware Description Language}
\newacronym{PIP}{PID}{Proportional Integral Derivative}
\newacronym{MCTS}{MCTS}{Monte-Carlo Tree Search}
\newacronym{DRL}{DRL}{Deep Neural-Network Reinforcement Learning}
\newacronym{LSTM}{LSTM}{Long Short-Term Memory}
\newacronym{EM}{EM}{Electron Microscope}
\newacronym{SoC}{SOC}{System on Chip}
\newacronym{CPU}{CPU}{Central Processing Unit}
\newacronym{RAM}{RAM}{Random Access Memory}
\newacronym{DSP}{DSP}{Digital Signal Processor}
\newacronym{LLRF}{LLRF}{Low Level RF}
\newacronym{VME}{VME}{Versa Module Eurocard}
\newacronym{LDRD}{LDRD}{Laboratory Directed Research and Development}
\newacronym{NAS}{NAS}{Neural Architecture Search}
\newacronym{FLOP}{FLOP}{floating-point operation}
\newacronym{SGD}{SGD}{Stochastic Gradient Descent}
\newacronym{QoS}{QoS}{Quality-of-Service}

\usepackage{graphicx}
\graphicspath{{images/}}
\usepackage[verbose,letterpaper,lmargin=1in,rmargin=1in,tmargin=1in,bmargin=1in]{geometry}

\usepackage{caption}
\usepackage{subcaption}
\usepackage{xcolor,colortbl}
\usepackage{multirow}
\usepackage{multicol}
\usepackage[compact]{titlesec}  
\titlespacing{\section}{2pt}{2pt}{1pt}

\usepackage{fancyhdr}
\usepackage{enumitem}
\usepackage{tabularx}
\usepackage[acronym]{glossaries}

\usepackage[colorlinks, urlcolor=blue, linkcolor=blue]{hyperref}

\renewcommand\thesection{\Alph{section}}
\renewcommand\thesubsection{\thesection.\alph{subsection}}

\begin{document}

%\maketitle  % pair `maketitle` with `title` code above if going that way...
% don't mix titlepage with `maketitle`
\begin{titlepage}
\thispagestyle{empty}

%\pagenumbering{roman}  % if needed, but watch doc style changes

%\section*{} 
\vspace{10pt}
\setlength{\parskip}{1em}

\noindent
{\Large Project title: \\
{\bf Accelerator Real-time Edge AI for Distributed Systems (\acrshort{READS})}}

\noindent
Institution: Fermi National Accelerator Laboratory (\acrshort{FNAL})

\noindent
Street address: Wilson Street \& Kirk Road, Batavia, IL 60510 \\
Postal Address: P.O. Box 500, Batavia, IL 60510

\noindent
Administrative Point of Contact: Hema~Ramamoorthi %, 630-840-6723, hema@fnal.gov

\noindent
PI: FNAL~Kiyomi Seiya %, 630-840-8187, kiyomi@fnal.gov \\
%Co-PI: FNAL~Nhan Tran, [Phone], ntran@fnal.gov \\
%Co-PI: FNAL~Vladimir Nagaslaev, [Phone], vnagasl@fnal.gov \\
%Co-PI: FNAL~Kyle Hazelwood, 630-840-5199, kjhi@fnal.gov \\
%Co-PI: FNAL~Michelle A Ibrahim, [Phone], cadornaa@fnal.gov \\
%Co-PI: FNAL~Brian Schupbach, 630-840-2293, schupbach@fnal.gov \\
\noindent
Submitting User Facility: Fermilab Accelerator Complex at FNAL

\noindent
DOE National Laboratory Announcement Number: \textbf{LAB 20-2261} \\
DOE/Office of Science Program Office: High Energy Physics (HEP) \\
DOE/Office of Science Program Office Technical Contact: Dr. John Boger

%%\noindent
%%PAMS Letter of Intent Tracking Number: LOI-00000???

\noindent
%Duration of project: 3 years

%\noindent
%Institutional Co-PIs:\\[8pt]
%\noindent
%Northwestern University: Seda Memik, Professor, seda@northwestern.edu, (847)491-7378\\
%\noindent
%Northwestern University: Han Liu, Associate Professor, hanliu@northwestern.edu, (847)491-2793\\

%\noindent
%Senior/Key Personnel:\\[10pt]
%\noindent
%Senior Person (Institution, email@example.com, (999)555-9999\\

% Students and postdocs, if named
%\noindent
%Additional Personnel:\\
%~\\

%\begin{tabular}{ | >{\raggedright}p{1.0in} |
   %>{\raggedright}p{0.8in} | >{\raggedright\arraybackslash}p{0.6in} |   >{\raggedright\arraybackslash}p{0.6in} | >{\raggedright\arraybackslash}p{0.6in} |
   %>{\raggedright\arraybackslash}p{0.6in} |}
%\hline
 %{\bf Names} &  {\bf Institution} &  {\bf Year 1 Budget} & {\bf Year 2 Budget} & {\bf Year 3 Budget} & {\bf Total Budget}\\
%\hline
%\hline
 %Kiyomi Seiya & FNAL     & 577,383 & 577,253 & 483,576 & 1,638,212 \\ \hline
%Seda Ogrenci-Memik  & Northwestern University &  210,512 &  219,789 &  227,029 & 657,330 \\ \hline
%\hline
%Totals          &          & 787,895 & 797,042 & 710,605 & 2,295,542 \\ \hline
%\end{tabular}

% {\color{red}
% To do list:
% \begin{itemize}[noitemsep,topsep=0pt]
%     \item citations, fix format (nhan)
%     \item biosketches, formatting and Accelerator LDRD funding (nhan)
%     \item current and pending (nhan)
%     \item letters of commitment (kiyomi, nhan)
%     \item fix timeline plot (kyle)
%     \item cover page (kiyomi)
%     \item abstract
% \end{itemize}
% }

\newpage

\end{titlepage}

\thispagestyle{empty}
\tableofcontents
%\listoffigures   % if needed -- generally don't want to waste space on this
%\listoftables    % if needed

% start page count after the title page and toc
\clearpage
\pagenumbering{arabic}

%\chapter{Introduction}

\section{Introduction and Motivation}

%\subsection*{AI for AD operation}
Over the last decade, \acrlong{ML} (\acrshort{ML}) technologies have slowly made their way into the accelerator community. Rapid advances in recent years in deep learning, particularly reinforcement learning for control system applications and the accessibility of deep learning in embedded hardware, have generated renewed interest and spawned a number of applications~\cite{ml}. 

\begin{figure}[tbh]
\centering
\vskip-15pt 
  %\fbox
  {\includegraphics[keepaspectratio, width=0.8\textwidth]{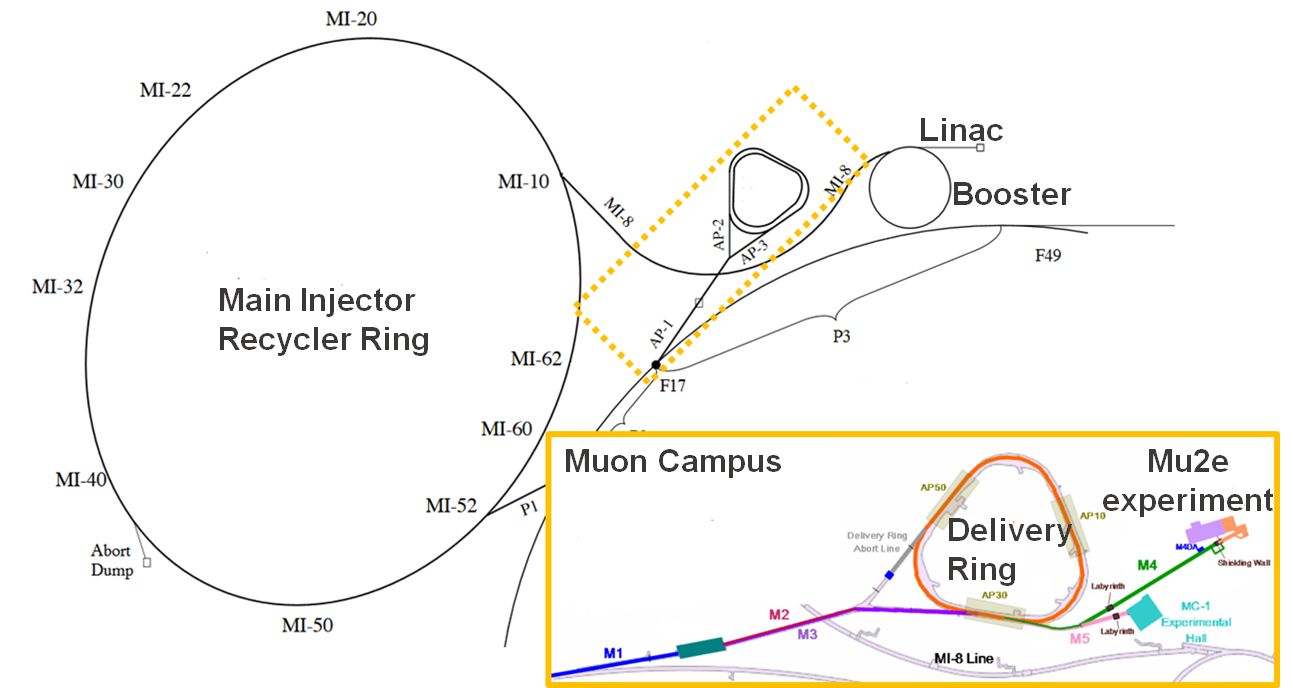}}
  \caption{Map of Fermilab accelerator complex with inset zoom of Muon Campus}
  \label{fig:muon_campus}
  \vskip-10pt 
\end{figure}

The Fermilab Accelerator Complex, shown in Fig.~\ref{fig:muon_campus}, has provided \acrlong{HEP} (\acrshort{HEP}) experiments with proton beams for nearly fifty years. %and has the potential to greatly benefit across many areas from modern machine learning techniques.
%In September 2011, the Tevatron collider ceased operations and 
The current focus of the laboratory is its world-class experimental program at the intensity frontier. 
%While increasing raw beam intensity certainly presents its own challenges, keeping losses to a minimum and maintaining beam stability turns out to be, in many ways, the main challenge. Achieving beam stability requires fine tuning and real-time parameter optimization. \acrshort{ML} methods are perfectly suited to tackle the challenge of learning the optimal management of a set of parameters with complex relations. Stepping beyond experience-based reasoning of human operators, which is inherently limited, and applying \arclong{ML} technologies are key to the success of future intensity upgrades. 
While increasing beam intensity certainly presents its own challenges, preserving beam size while minimizing beam losses -- particles lost through interactions with the beam vacuum pipe -- turns out to be, in many ways, the main challenge.
The accelerator is controlled via a complex system of hundreds of thousands of devices.
Enabling fine tuning and real-time optimization of their parameters using \acrshort{ML} methods and stepping beyond experience-based reasoning of human operators are key to the success of future intensity upgrades. 

Our objective will be to integrate \acrshort{ML} into accelerator operations and furthermore, provide an accessible framework, which can also be used by a broad range of other accelerator systems with dynamic tuning needs. 
%make the technology as accessible as possible. 
To successfully maximize the benefits of applying \acrshort{ML}, we will consider the following:

\textbf{{Real-time edge \acrshort{ML} system optimization}}:
An accelerator involves a complicated array of regulation loops for power supplies, \acrshort{RF} and other control systems. The gains of the regulation loops are manually optimized and fixed for operations. %The gains of these systems are traditionally optimized for a fixed set of a prior operating conditions. 
In reality, beam distribution and intensity are dynamic quantities that evolve during acceleration.  Consequently, %and in the ideal case, 
these dynamic systems should ideally re-examine operating conditions in near real time. 
%The possibility of which 
This requires an \acrshort{ML} model capable of reacting to changes in the system on a sufficiently short, milliseconds, time scale. %In addition, power supplies and other control systems have been replaced, upgraded and adapted to the existing complex over time. Developing \acrshort{ML} systems capable of handling the diversity of existing systems presents an additional challenge.

\textbf{{Fast, intelligent distributed systems}}:
%While most control systems are localized, from extensive experience, operators and machine experts know that they are not orthogonal. 
Due to the large physical scale of particle accelerators, control systems tend to be spread across the facility. Optimizing the performance of each machine as well as the overall performance of the complex, therefore implies a fast data transfer system allowing for real-time communication between subsystems, machines, and computer resources tasked with running the ML algorithms. 

Our project, {\bf Accelerator READS} will develop \acrshort{ML} methods and their edge implementation within large scale accelerator systems. Fermilab is a leader in the development of real-time embedded edge \acrshort{ML} devices
for system control and has leveraged \acrshort{ML} to improve the efficiency and accuracy of HEP experiments such as the \acrlong{CMS} (\acrshort{CMS}) experiment~\cite{Duarte:2018ite}. Using the internal \acrlong{LDRD} (\acrshort{LDRD}) program, Fermilab has demonstrated that a single \acrshort{ML} system can improve accelerator performance.
However, connecting embedded \acrshort{ML} systems together to analyze and control multiple complex structures in coordination has not been done. 
The application of this technology to accelerators would be an evolution in capabilities towards fast, distributed, and high-performance control and operations of the Fermilab accelerator facility.

Methods and tools resulting from Accelerator READS will be relevant for design of a variety of complex and distributed controllers. We will demonstrate the effectiveness of our proposal with two experiments of significance: the \acrshort{Mu2e} spill regulation system and the de-blending of the Main Injector (MI) and Recycler Ring (RR) beam losses. 

\subsection*{\acrshort{Mu2e} Experiment and Schedule}
    
The \acrlong{Mu2e} (\acrshort{Mu2e}) is one of the major experiments in the Fermilab program whose construction phase is nearing completion. It is expected to come online in March 2022, succeeding the g-2 experiment. \acrshort{Mu2e} will search for ultra-rare lepton flavor violating muon to electron direct conversion with a sensitivity four orders of magnitude higher than any previous attempts~\cite{Bartoszek:2014mya}. To achieve the goal, the experiment uses sophisticated apparatus and imposes very strict requirements on the quality of the proton beam delivered to the production target.

The former Antiproton Source storage rings and transport lines have been re-purposed and upgraded into what is now known as the Muon Campus, shown in Fig. 1. The 400 MeV protons are accelerated to 8 GeV in the Booster, transferred into the Recycler Ring where the beam bunch structure is optimized before being finally injected into the \acrlong{DR} (\acrshort{DR}). While the bunch is circulating in the \acrshort{DR}, it is gradually extracted to the target which is referred to as a spill.

The Mu2e experiment requires that $3.6 \times 10^{20}$ Protons on Target (POT) be delivered over a running period of 3 years.  Minimizing the downtime of the accelerator complex will be key to meeting this requirement as soon as possible. The experiment also requires that the extracted beam intensity within a spill be very uniform and the beam losses remain below 2\% of the total beam power in order to control radiation and thereby reducing the equipment activation and personnel exposure.  
\subsection*{Employing Machine Learning}

 We aim to use \acrlong{ML} techniques to improve overall delivered beam performance to the \acrshort{Mu2e} experiment and boost physics output by\:
\begin{itemize}[noitemsep,topsep=0pt]
    \item improving the real-time spill regulation 
    with the use of reinforcement learning algorithms for guided operations optimization thereby, increasing the \acrlong{SDF} of slow spill extraction; algorithms will be developed with the aid of a digital twin of the spill regulation system
    \item reducing beam aborts with intelligent and semi-autonomous operations by deploying de-blending and de-noising techniques to decouple overlapping beam losses in the \acrlong{MI} enclosure, thereby, increasing the overall uptime of the \acrlong{RR} (see Fig.~\ref{fig:muon_campus}), as well as delivering the beam to many other experiments across the entire complex 
\end{itemize}
Both tasks will leverage capabilities unique to Fermilab in implementing real-time \acrlong{ML} algorithms in embedded systems with millisecond scale feedback. Accelerator READS will develop shared machine learning tools, system instrumentation, and algorithmic techniques which can be deployed for beam delivery systems like \acrlong{PIP-II} (\acrshort{PIP-II}) for future neutrino experiments and across other large accelerator complexes. 

\section{Proposed Research and Methods}
% {\bf Identify the hypotheses to be tested (if any) and details of the methods to be used including the integration of experiments with theoretical and computational research efforts.}\\

% {\color{red} Prose around high-level inputs -NT}

% Explain the overall scope and the reason to transmit information across the laboratory.
% Detail each of the steps:
% \begin{itemize}
%     \item extract low latency signals from the RWMs, BLMs, quads, RFKOs at high rate
%     \item transmit them to a centralized board using common ethernet protocol and to data storage element in SCD
%     \item Provide compute infrastructure for O(100Tb) datasets
%     \item Build ML models, surrogate models and online agent, for controls
%     \item Implement ML model on FPGA and setup online training system 
% \end{itemize}

A collaboration between Fermilab and Northwestern University will pull together the talents and resources of accelerator physicists, beam instrumentation engineers, embedded system architects, \acrlong{FPGA} design experts, and \acrshort{ML} experts to solve complex real-time accelerator controls challenges which will enhance the physics program. Deploying \acrlong{AI} (\acrshort{AI}) systems at the millisecond scale or faster for low latency processes is beyond traditional online intelligent processing systems or human-in-the-loop controls. Developing the convergent aspects of advanced embedded system development
and \acrshort{ML} applications will build new \acrshort{AI} capabilities that can be leveraged across Fermilab and resolve similar challenges at existing and future \acrshort{DOE} facilities.

There are several elements to this proposed research deploying a range of techniques. In Section~\ref{sec:srs} and Section~\ref{sec:blms}, we describe two important challenges for delivering quality beam to the \acrshort{Mu2e} experiment: the spill extraction of the proton beam from the Muon Campus \acrlong{DR} to the \acrshort{Mu2e} experiment and the disentangling of beam loss signatures from the \acrshort{MI} and \acrshort{RR} accelerators in order to increase uptime of beam delivered to the \acrshort{Mu2e} experiment, and consequently also improving overall performance of the Fermilab accelerator complex. In these sections, we will discuss how to augment beam monitoring instrumentation for real-time \acrshort{ML} capabilities and where \acrshort{ML} can be used to enhance the systems. In Section~\ref{sec:pipeline}, we discuss how to aggregate the signals from distributed instrumentation into a single data pipeline for online controls systems and for offline storage to perform data analysis and train \acrshort{ML} models.  Section~\ref{sec:mlmodels} will reformulate the accelerator challenges into the \acrshort{ML} domain and discuss how modern machine learning techniques will be deployed for these applications.  Finally, in Sec.~\ref{sec:mlsystem}, we will discuss implementation and system aspects of deploying machine learning models within the operational controls system of the Fermilab accelerator complex and the novel challenges for building intelligent embedded hardware systems.  \textbf{PI Seiya}, who has 20 years of experience in Fermilab AD operation and beam physics, will be coordinating activities across the entire project.

%%%%%%%%%%%%%%%%%%%%%%%%%%%%%%%%%%%%%%%%%%%%%%%%%%%%%%%%%%%%%%%%%%%%%%%%%%%

\subsection{Regulation Loop for Mu2e Slow Spill} 
\label{sec:srs}

%For this aspect of the project, the goal
The goal of the regulation loop is to deliver a spill with a highly stable intensity profile. This necessitates fine control of the spill regulation process. %Our goal is 
We propose to deploy an \acrshort{ML} agent to adjust the \acrlong{SRS} (\acrshort{SRS}) parameters in real-time by providing feedback at the approximately milliseconds timescale. \textbf{Co-PI Nagaslaev} and \textbf{Co-PI Ibrahim} are project leads in the spill regulation system design for the \acrshort{DR} and will guide \acrshort{ML} integration.  

\subsubsection*{Operational Overview for Slow Extraction}

%The full beam delivery accelerator chain is shown in Fig.~\ref{fig:muon_campus}. The former Antiproton Source storage rings and transport lines have been re-purposed and upgraded into what is now known as the Muon Campus (zoomed in). Currently the Muon Campus is serving the beam for the g-2 experiment. Once g-2 is completed, the Muon Campus will start delivering beam to the Mu2e experiment.

%\acrshort{Mu2e} uses 8~kW of 8~GeV protons from the Booster. Two batches of approximately $4\times 10^{12}$ protons are sent to the \acrlong{RR} (\acrshort{RR}) for each 1.333~sec \acrlong{MI} (\acrshort{MI}) cycle. In the \acrshort{RR}, the beam is then re-bunched into 8 bunches, and each will be individually transferred to the \acrshort{DR}. Once a bunch of $1\times 10^{12}$ protons is injected into the \acrshort{DR}, it is then slowly extracted to the Mu2e production target within a 43.1~msec spill period. 

The layout of the slow extraction components in the \acrshort{DR} is shown in Fig.~\ref{fig:pid_loop}. 
%The third integer resonant extraction condition in the \acrshort{DR} is established by exciting two families of the harmonic sextupole magnets. The tune ramp quad magnets drive ("squeeze") the machine operating point to the exact resonance, gradually pushing the circulating beam into the resonance stop band. 
The process of slow beam extraction is achieved by creating the resonance condition with the use of dedicated sextupole magnets. The fraction of beam particles that fall into the resonance and become unstable is controlled by the tune ramping quads. These magnets drive (`squeeze') the machine operating point (tune) to the exact resonance, gradually pushing the entire circulating beam out of the stable condition.
%As unstable particles in the stop band drift towards 
As unstable particles drift towards
the machine aperture, they get intercepted in the \acrlong{ESS} (\acrshort{ESS}) and deflected towards the extraction line. The extracted beam is a stream of $\sim$200~nanoseconds. wide pulses, separated by the \acrshort{DR} revolution period of 1,695~nanoseconds. 

\subsubsection*{Spill Regulation System}

The objective of the \acrshort{SRS} is to maintain the intensity uniformity of the pulses extracted from the \acrshort{DR} to the target area. The quantitative metric which defines beam stability is called the \textit{\acrlong{SDF}} (\acrshort{SDF}) which is defined as:
\begin{equation}
    \text{SDF} = \frac{\langle I \rangle^2}{\langle I^2 \rangle}
    \label{eq:sdf}
\end{equation}
Here, \emph{I} is a single pulse intensity. The design value of the \acrshort{SDF} is 60\%. It is very important for the experiment to have this value as high as possible. Large intensity variations in the spill will saturate the throughput capability of the Mu2e data acquisition system and create dead time issues in the readout system resulting in the loss of useful detector data.  
Substantial increases of the \acrshort{SDF} value typically require a large effort over a long period of time \cite{sdf_c}, \cite{sdf_j}. This proposal has a great potential to significantly improve on the \acrshort{SDF} limit through the exploration of ML and hence, increase the Mu2e uptime.

\begin{figure}[tbh]
\centering
  %\fbox
  {\includegraphics[keepaspectratio, width=0.9\textwidth]{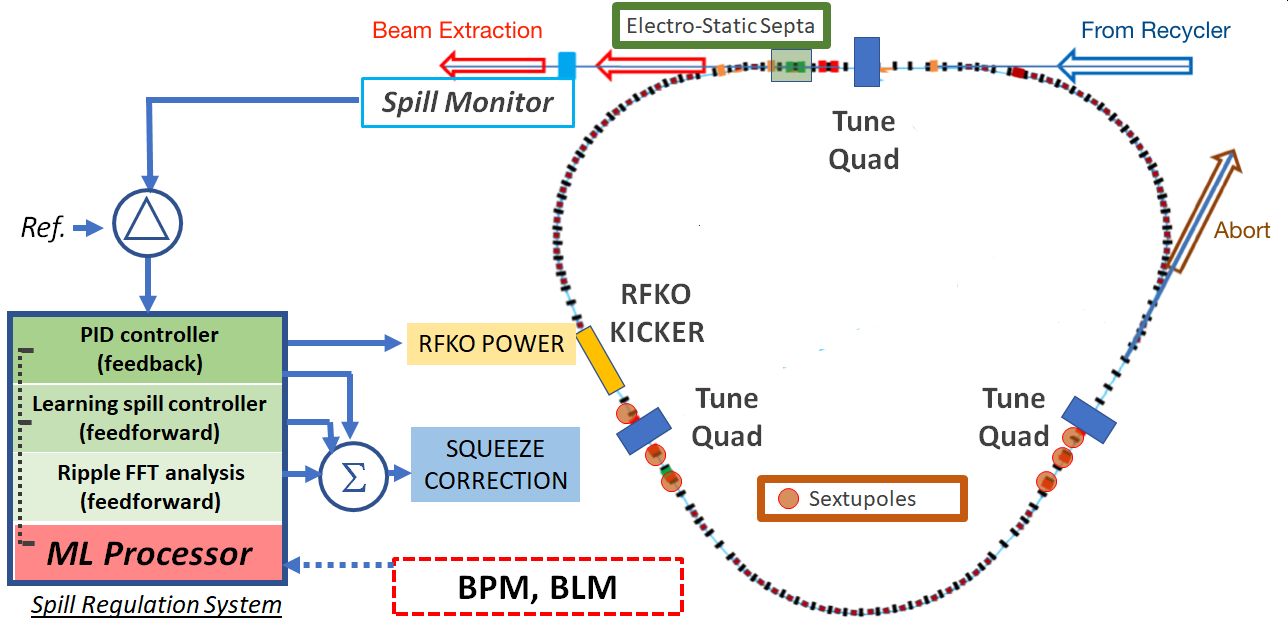}}
  \caption{Delivery Ring and control loop for the Spill Regulation System}
  \label{fig:pid_loop}
  \vskip-10pt 
\end{figure}

The \acrlong{SRS} is implemented in the Intel Arria10 \acrshort{SoC} with custom carrier board (Fig.~\ref{fig:pid_loop}, left) and it has several control loops. 
The main way to control the extraction rate is by regulating the voltage reference current to drive the tune ramping quads. The ramping curve of quad excitation determines the shape of the beam intensity profile during the spill. 
The other handle to control the extraction rate is the fast modulation of the \acrshort{RFKO} system.  \acrshort{RFKO} is used to effectively heat the beam in the horizontal plane and accelerate the diffusion of particles into the unstable areas. 
The \acrshort{SRS} uses the \acrshort{PID} control loop to simultaneously regulate these two primary beam correction elements.

The loop will use the spill monitor intensity measurements to monitor the instantaneous and integrated spill parameters. This signal will be provided as a reference for the fast \acrshort{PID} loop and as an input for the two other loops.

\subsubsection*{\acrshort{ML} for the Spill Regulation System}
The \acrshort{PID} loop parameters (gains) need to be re-optimized in real time. This can be done very effectively with \acrshort{ML}. Moreover, in this case  the loop gains can be expanded into a nonlinear time-dependent series to provide a better coverage of the regulation frequency range.

Bringing the ML agent into the process opens the way to extend the operational functions of the SRS with inclusion of new environmental inputs: (1) the turn-by-turn beam position monitor (BPM) signal and (2) \acrlong{BLM} (\acrshort{BLM}) in the \acrshort{DR}.

\begin{figure}[tbh]
\centering
  %\fbox
  {\includegraphics[keepaspectratio, width=0.45\textwidth]{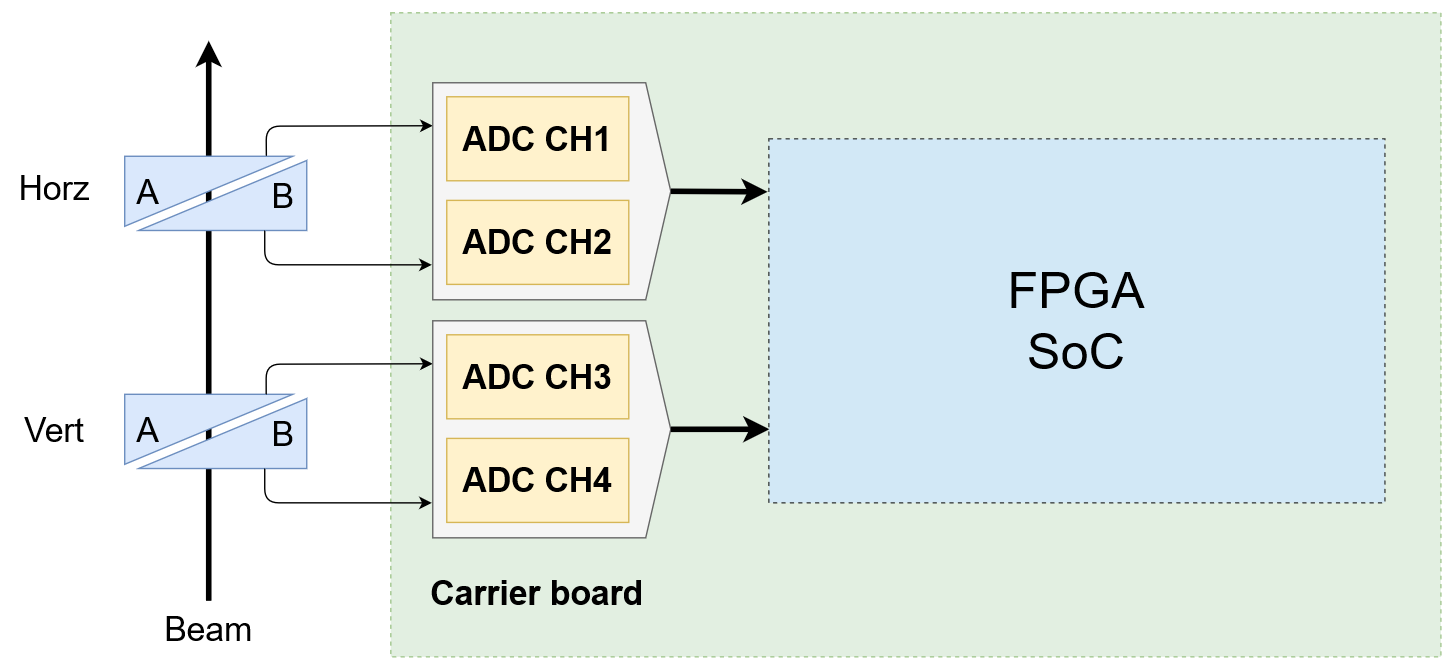}}
  \hspace{0.1in}
  {\includegraphics[keepaspectratio, width=0.45\textwidth]{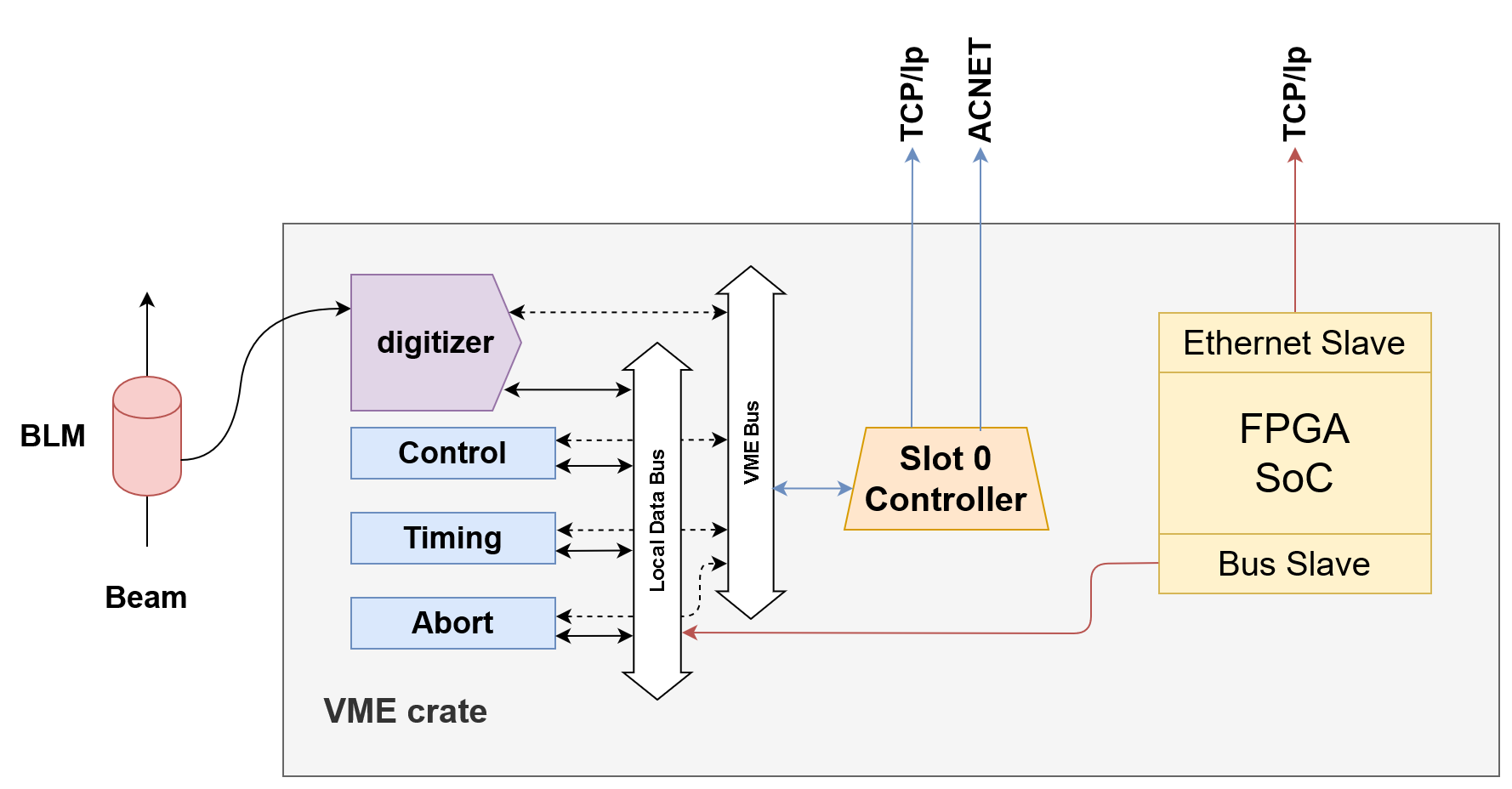}}
  \caption{Left: Readout for Turn-by-Turn BPM; Right: Beam Loss Monitor data extraction}
  \label{fig:soc_vme}
%   \vskip-10pt 
\end{figure}

The turn-by-turn beam position measurement (Fig.~\ref{fig:soc_vme}-left) can be used to calculate the injection beam steering error. 
This error may have a substantial random component, leading to an unpredictable distortion of the beam shape and therefore, the spill profile. The analog signal trace for the first 50 turns can be digitized and analyzed in the \acrlong{SRS} to provide immediate information on the beam shape change for every spill. The \acrshort{ML} process will help to determine the algorithm for
spill-to-spill corrections for the squeeze waveform.   

Small beam losses at the design level near 2\% create significant radiation constraints with the beam power of 8~kW. To mitigate this, 1~m thick iron shielding is installed above the \acrshort{ESS} in the tunnel. Safety systems are in place to instantly shut down the beam operations if beam losses exceed the permitted level. Monitoring and improving the losses in real time becomes now possible with the use of the \acrshort{BLM} data (Fig.~\ref{fig:soc_vme}-right). The data analysis is very similar to that proposed for the de-blending of the RR/MI losses  and will be discussed in detail in section~\ref{sec:blms}. The \acrshort{ML} process will track the changes in the loss pattern to identify new sources of beam loss and 
initiate corrections to the extraction control elements. This will improve the uptime in the DR.

%%%%%%%%%%%%%%%%%%%%%%%%%%%%%%%%%%%%%%%%%%%%%%%%%%%%%%%%%%%%%%%%%%%%%%

\subsection{De-blending of Main Injector and Recycler Beam Losses}
\label{sec:blms}
The second demonstration of our Accelerator \acrshort{READS} concept will be through the de-blending system. \textbf{Co-PI Hazelwood} is a senior engineering physicist in the Main Injector department at Fermilab and has extensive experience in accelerator controls and has designed \acrshort{ML} algorithms for beam physics reconstruction.  
\subsubsection*{Operational Overview of Beam Loss Monitoring in the Main Injector Enclosure}
The Main Injector enclosure houses two accelerators; the Main Injector, which is a 120~GeV conventional powered magnet synchrotron and the Recycler, which is an 8~GeV permanent magnet ring. The Recycler was originally commissioned as an anti-proton storage ring for the Tevatron collider. Anti-proton intensities in the Recycler during the collider era were relatively small compared to the proton intensities going through the Main Injector, hence, de-coupling beam loss from the machines was of little concern. Since the end of the Tevatron collider era, the Recycler has been re-purposed as a proton stacker for 120~GeV \acrshort{NuMI} beam operation \cite{Ainsworth:2018ntr} as well as 8~GeV muon campus beam operation~\cite{Bartoszek:2014mya}. The new modes of operation mean high intensity proton beams are often simultaneously running in both the Recycler and the Main Injector. High intensity beam in the Recycler also means the possibility of higher beam loss. Beam losses in the Recycler often rival those of its sibling machine, the Main Injector. Beam losses in the Main Injector enclosure are monitored for both tuning the accelerators and machine protection. Seven beam loss monitor nodes are distributed around the 2.2~mile Main Injector enclosure (Fig.~\ref{fig:muon_campus}) to monitor all 259 operational ionization chamber beam loss monitors. Readings from these nodes are transmitted to \acrshort{ACNET}, Fermilab's accelerator control system. The nodes also actively compare loss readings against set thresholds to trigger beam aborts for either machine. While separate loss thresholds may be set for each machine, in practice the limits for both machines are set conservatively, because we are unable to accurately attribute loss to the correct machine. This means that the vast majority of the time losses from one machine will cause a beam abort in both machines resulting in unnecessary downtime of the sibling machine.

\subsubsection*{Expert Beam Loss Machine Attribution}
\begin{figure}[tbh]
\centering
 %\fbox
 {\includegraphics[keepaspectratio, width=0.45\textwidth]{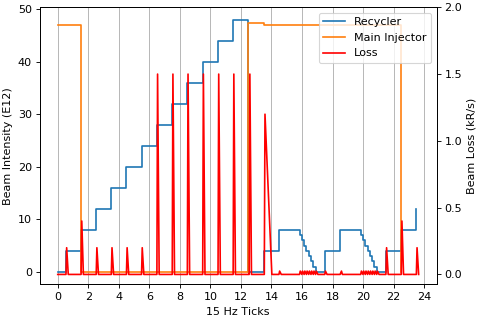}}
 \hspace{0.1in}
 {\includegraphics[keepaspectratio, width=0.45\textwidth]{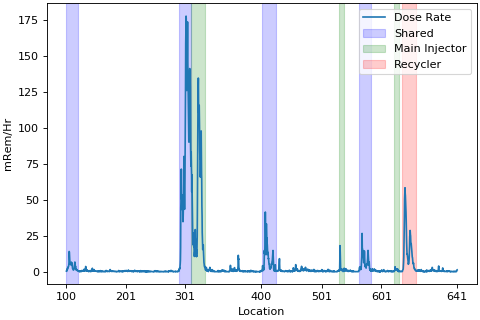}}
 \caption{Left: Time dependency of MI/RR beam losses, Right: Location dependency of MI/RR beam loss as seen from tunnel dose rates}
 \label{fig:loss_dependency}
  \vskip-10pt 
\end{figure}

% While no automatic way of de-blending losses between \acrlong{MI} and \acrlong{RR} currently exists, machine experts can often distinguish losses between machines. 
While losses are currently attributed to a machine based on timing, this method alone is insufficient and often inaccurate. Machine experts can often distinguish losses between machines better than the current system. Loss patterns exist such that those with knowledge of the machine can look at all the \acrshort{BLM}s around the enclosure and pick out patterns in time and location to help attribute loss to a particular machine. As Fig.\ref{fig:loss_dependency}-left shows, time of loss can be a great indicator as to what machine caused the loss. Examples of time based losses are start of ramps, collimation bumps, injection and extractions, all of which are particular to a cycle and indicate a certain machine. Location and shape of loss patterns are also very helpful. The \acrlong{RR} and the \acrlong{MI} have known aperture restrictions, that are areas where the beam is most likely to be lost when problems arise. Some of the Main Injector's and Recycler's aperture restrictions are unique to their respective machine while some locations have overlapping common loss points (Fig.\ref{fig:loss_dependency}-right). In places where the two machines share aperture restrictions near the same physical location, such as injection and extraction locations, the loss pattern shape can help discern which machine caused the loss. Losses often appear first in one  \acrshort{BLM} followed by a "spray" of losses. Depending on the machine, loss spray may be revealed or masked by components in the accelerator. The goal of this project would be to replicate and improve upon what experts are attempting to a limited extent to attribute loss to a machine.

% \begin{figure}[tbh]
%     \centering
%     %\fbox
%     \includegraphics[keepaspectratio, width=0.9\textwidth]{LAB_20-2261_Embedded/images/doe_foa_ml_2020_loss_time_dependency_wide.png}
%     \caption{Time dependency of Main Injector and Recycler beam losses.}
%     \label{fig:loss_dependency}
% \end{figure}

\subsubsection*{Real-time Beam Loss De-blending using Machine Learning}
\begin{figure}[tbh]
\centering
  \includegraphics[keepaspectratio, width=0.8\textwidth]{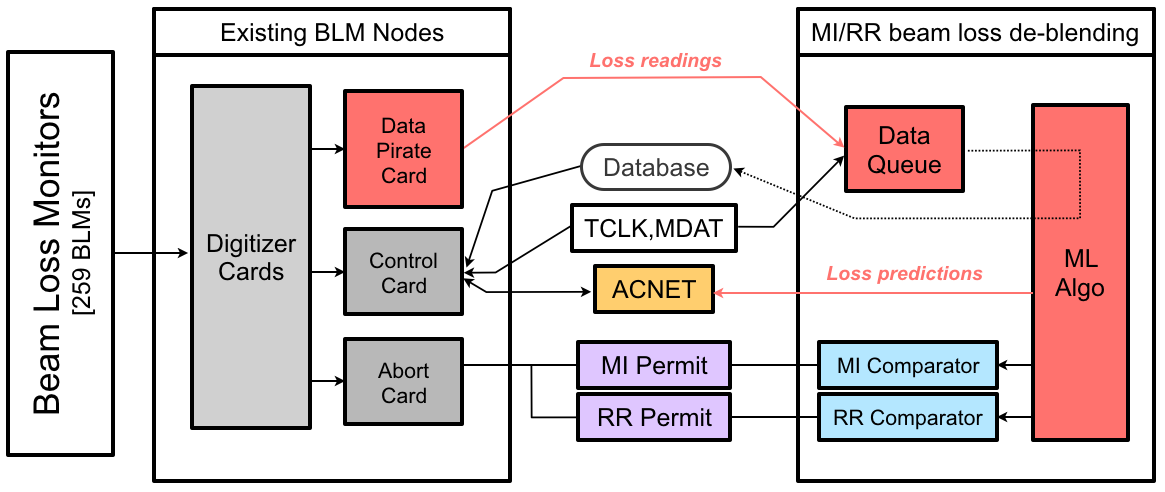}
  \caption{Main Injector and Recycler beam loss de-blending conceptual design}
  \label{fig:loss_deblending_flow}
\end{figure}

It is beyond the scope of this proposal to completely replace the existing Main Injector enclosure beam loss monitoring system. The cost of doing so is too great and current schedules require that the \acrshort{BLM} system remains functioning. For these reasons, we are proposing a system that leaves the existing \acrshort{BLM} infrastructure virtually unchanged. Instead, the current \acrshort{BLM} nodes would be utilized to feed beam loss readings to a central node that will then use a trained \acrshort{ML} model to de-blend losses and label them by machine. As seen in Fig.~\ref{fig:loss_deblending_flow}, to facilitate reading and transmitting loss 
%readings 
measurements at high rates (200-1440 Hz) without tasking the existing \acrshort{BLM} nodes, new loss reading ``pirate" cards will be developed to listen to the control bus on each crate and transmit readings over ethernet to the central node. The central node would use a data queue to aggregate loss readings from around the ring as well as information deemed to be useful such as clock or beam intensity. Data sets created from the data queue would then be fed to the \acrshort{ML} model. The output from the \acrshort{ML} model will be made available to \acrshort{ACNET} for machine tuning and diagnostics. The \acrshort{ML} output will also be used to decide whether or not to disable a machine's beam permit due to excessive beam losses. In order to properly protect the machines and serve as a useful tool for tuning, ideally the ML model should provide output at no less than 200 Hz frequency.

%%%%%%%%%%%%%%%%%%%%%%%%%%%%%%%%%%%%%%%%%%%%%%%%%%%%%%%%%%%%%%%%%%

\subsection{Data Pipeline, Storage, and Computation}
\label{sec:pipeline}

Successfully implementing \acrshort{ML} in an embedded systems environment will require additional hardware and software to support a data processing pipeline which is capable of ingesting, archiving, validating, and training an appropriate set of accelerator data. 
%In this section, we will discuss how the data will be preserved, shared, validated, and processed. 
We consider three primary phases of the data flow: (i) the data being created at the beam instrumentation after digitization and transmitted to the online processing \acrshort{ML} algorithm; (ii) nearby edge compute resources for re-optimizing the agent in real-time; and (iii) offline data storage and archiving for large scale training using on-premises resources or the cloud.  

For the phases (i) and (ii) of the data flow, we require communication protocols for streaming digitized data off of custom data buses to \acrshort{FPGA}s 
%(Field Programmable Gate Arrays) 
capable of handling large data rates with low latency.  The data ingested by the \acrshort{FPGA} is then sent to a more traditional CPU architecture where it will be transmitted by ethernet to the processing ML algorithm, which in turn will be deployed on another \acrshort{FPGA} device.  This fast generic FPGA-CPU communication will be a central capability of the project and implemented for several data pipelines, transmitting sensor data and also streaming data from the online algorithm to offline computing elements.  

Fermilab \acrlong{AD} has been developing control modules based on the Intel Arria10 \acrlong{SoC} (\acrshort{SoC}), an integrated ARM-based CPU+\acrshort{FPGA} device, and is planning to standardize them for control systems and instrumentation. The two projects described in Section~\ref{sec:srs} and Section~\ref{sec:blms}, require development of \acrshort{FPGA} modules which allow for the reception of data packets from the existing \acrlong{VME} (\acrshort{VME}) crate and transmittance to a central module at a high speed. %Development of this board will complete by February 2021.
We envision that this board will be the common development platform for the proposal
%for this project 
and has the requisite capabilities to deliver on all the needs of this project.

Phase (iii) of the data flow requires storing and archiving data from the online processing node to build a large scale digital twin of the accelerator system.  
Commodity disk space, time-series databases, cloud services for the large-scale phases of training, and on-site compute nodes (CPUs) with \acrshort{GPU} support will all be a part of the available data ecosystem. These resources will help with creating beam simulations for initial model training, data pre-processing, and organization for beam data itself. For model training, we will consider on-premises GPUs for smaller scale and simple \acrshort{ML} model training, while for large-scale burst training, we will allot cloud-scale resources.  

\begin{figure}[tbh]
\centering
  \includegraphics[keepaspectratio, width=0.95\textwidth]{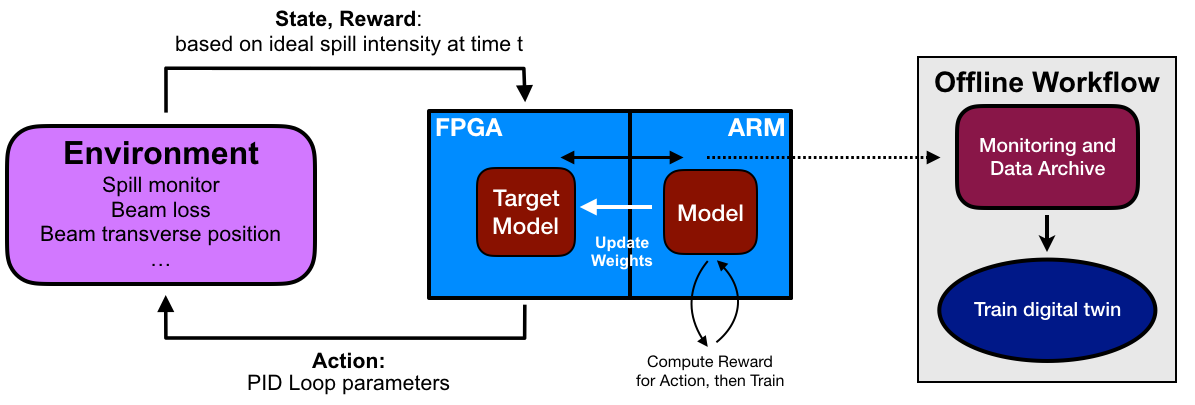}
  \caption{A reinforcement learning schematic for Mu2e spill extraction application including data flow pathways across the control system}
  \label{fig:rl}
\end{figure}

An illustration of an example of the entire data pipeline with ML model feedback loop for the spill regulation system is presented in Fig.~\ref{fig:rl}.  It is also relevant for the beam loss de-blending application as well. The illustration shows the multiple data pathway phases from the raw instrumentation data transferred to the ML processing, the FPGA to CPU communication, and the streaming to offline data storage for large \acrshort{ML} training workflows. 

In the next section, Section.~\ref{sec:mlmodels}, we will discuss the aspects of Fig.~\ref{fig:rl} pertaining to the development of the machine learning models and the creation of input and training data.  In Section~\ref{sec:mlsystem}, we will discuss how we will implement those \acrshort{ML} models into the \acrshort{FPGA} fabric itself for real-time online operations.  

%%%%%%%%%%%%%%%%%%%%%%%%%%%%%%%%%%%%%%%%%%%%%%%%%%%%%%%%%%%%%%%%

\subsection{Machine Learning Model Development}
\label{sec:mlmodels}
In this section we will elaborate on our proposed \acrshort{ML} methods that are tailored to the needs of different accelerator control problems. This part of the project will be led by \textbf{Co-PI Liu} who has an extensive expertise and background in \acrshort{ML} and leading \acrshort{ML} projects.

\subsubsection*{Reinforcement Learning for %Controls for Mu2e slow spill} 
Controlling the Spill Regulation System}
\paragraph*{\acrshort{SRS} simulations}
A full suite of physics simulations can be used to study the regulation algorithms %with addition of 
enhanced by \acrshort{ML} while the beam instrumentation and controls are being developed. This will inform the early stages of the \acrshort{ML} model studies in a well-understood simulation environment.   

The detailed simulations of the slow %extraction 
spill regulation have to include a large number of physics processes and the statistical nature of the extraction process.
%the entire sequence of processes: generation of spill intensity variations, the beam transport in the accelerator, the extraction process, the spill monitor response, signal processing in the \acrshort{SRS}, \acrshort{SRS} output and the action of the beam regulation devices. By far, the most consuming part in this loop is the beam transport, as it has to track hundreds of thousands of macro-particles through hundreds of field shaping elements in the storage ring, including nonlinear elements forming the resonance – for the duration of many thousands of revolutions in the machine. 
Such a modeling of the extraction process takes substantial computing time using grid resources. We have expertise on performing 
%a good experience with these 
simulations at this complexity~\cite{dr_re}.
For the purposes of studying the regulation process, the model can be significantly simplified.
The new model would replace the most time-consuming part of beam dynamics with an analytical model, which is, with a few exceptions, still adequately representative for the most significant extraction response to the sources of variation. This model can be used for fast MATLAB simulations to test the real-time firmware and provide data sets for the offline training of \acrshort{ML} models.
%offline training. 
%board in real time and provide the data sets for the ML offline training. The FPGA algorithms developed at this stage will provide an input for defining the SRS design specifications to accommodate the ML capabilities on the board.

\paragraph*{ML Model Building}  The primary function of the control loop in the \acrshort{Mu2e} slow spill is to tune the \acrshort{RFKO} power and
%the tune 
the quadrupole correction currents.  The goal is to maintain a consistent spill intensity to achieve a high \acrlong{SDF} as defined in Eq.~\ref{eq:sdf}. We plan to %apply machine learning techniques to improve the performance of the control loop.  Specifically, we 
target a class of reinforcement learning techniques \cite{bertsekas1996neuro} which model the control loop as an online \textbf{\textit{agent}} taking \textit{\textbf{action}} to tune the \acrshort{RFKO} and quadrupole systems. The action $a_t$, which will be selected by the learned policy $\pi$ based on the current state $s_t$, generates a stimuli to the \textit{\textbf{environment}} which updates the state of the \textit{\textbf{agent}} and a \textit{\textbf{reward}} $r_t$ is computed based on the ideal spill intensity for a 100\% \acrlong{SDF}.
This reinforcement learning loop is depicted in Fig.~\ref{fig:rl}. Our main goal is to maximize $\sum_{t} r_t$.

The online agent takes a large number of environmental inputs of various timescales. To reduce system risk and fully harness the potential of the reinforcement learning agent against a traditional \acrshort{PID} controller, we propose a phased implementation of the model considering two paradigms: model-free and model-based learning. Consider a traditional \acrshort{PID} controller based on a model $\mathcal{M}(\alpha,\beta,\gamma,...)$ where $\alpha,\beta,\gamma$ are model parameters.  In a model-based paradigm, the online learning agent learns the parameters of $\mathcal{M}$ and exploits them to evaluate the transition probability from state $s$ to state $s'$ if action $a$ is performed.  
The development of the digital twin is performed offline using a large training sample. We will fine tune this model with meta learning techniques to better imitate the real environment.

One striking feature of our proposed machine learning algorithms is their real-time performance guarantee. We propose two strategies to 
%solve 
achieve this: (i) a hierarchical architecture; (ii) a \acrlong{QoS} (\acrshort{QoS}) protocol. Hierarchical reinforcement learning \cite{barto2003recent} incorporates hierarchies into the representation of actions taken by the online agent. Such a hierarchical representation eliminates unnecessary decision branches and sketches appropriate actions with lower granularity, thus increasing its time efficiency. In contrast, \acrshort{QoS} provides a systematic framework to trade off the decision speed and quality. An integration of the two delivers a stronger performance guarantee for the %obtained
resulting algorithms. 

\subsubsection*{Supervised Learning for De-blending Main Injector and Recycler Losses} 
%Training input size [~275 (259 blm readings, 12+ additional readings), 1] array of normalized floats.
%Training output size [259, 2], one hot labeled data, two classes [Recycler, Main Injector].
%Output size [259, 2] soft-max binary classifiers, two classes [Recycler, Main Injector]. Multiply model output (0.0 to 1.0) to original loss readings for each machine. 

\paragraph*{Creation of Machine Labeled Beam Loss Training Data} 
Current modes of operation do not allow much time to collect training data when the beam is only in one machine at a time. Creating a sufficiently large set of %enough 
accurate machine labeled beam loss training data by monitoring operations is not feasible. Accelerator beam loss simulations of the machines are also not very useful because losses often occur due to small imperfections such as beam pipe welds or miss-alignments that are not accounted for in our accelerator models. 
Therefore, an initial training data sample with dedicated beam studies using actual accelerator data will be collected during beam commissioning periods. In such a study, the beam would be injected into only one machine at a time and loss would be created in various ways at varying locations and times. All \acrshort{BLM}s would be recorded at a frequency needed to train a model for real-time loss de-blending. Other information such as clock (\acrshort{TCLK}), permit status, and \acrlong{MDAT} (\acrshort{MDAT}) would be recorded as well. 
%It may be that the existing \acrshort{BLM} system is in-capable of transmitting readings for all the \acrshort{BLM}s at a high enough frequency to create our training set. If so, the installation of the reading ``pirate" cards would be required before any training data could be created. 
Running conditions and operational modes in both the Main Injector and the Recycler change throughout the year. This project lends itself nicely to incremental training where data from operations is used to improve the initial \acrshort{ML} model. The use of incremental training will be explored to lessen the need for costly beam loss studies.

\paragraph*{ML Model Building} The goal of de-blending Main Injector enclosure losses is to 
%recognize 
differentiate between the beam loss from the Main Injector and the beam loss from the Recycler by analyzing the beam loss monitors' readings. We plan to implement a supervised machine learning algorithm to tackle this. 
In this problem, we have response variables $Y$ and inputs (predictors) $X$. $Y$ denotes the respective desired labels for categorizing the $k$ \acrshort{BLM}s, indicating whether the beam loss comes from the Recycler or the Main Injector. The inputs $X$, on the other hand, represent the readings from the \acrshort{BLM}s and other monitors. We aim to learn a prediction function that maps the inputs $X$ to the response variables $Y$ by minimizing a cross-entropy loss function to predict the true beam loss categories.   
%We aim to learn a prediction function $f$, where $\hat{Y}=f(X)$ and $\hat{Y}_k=P(Y_k=1|X)$, to minimize $\ell(f, X, Y) + \lambda R(f)$ where $\ell$ is the cross-entropy loss function and $R(f)$ i
%In this problem, we have response variables $Y$ and predictors $X$. $Y$ denotes the respective desired labels for the $k$ \arcshort{BLM}s, indicating whether the beam loss comes from the Recycler or the Main Injector. The predictors $X$, on the other hand, represent the readings from the \arcshort{BLM}s and other monitors. We aim to learn a prediction function $f$, where $\hat{Y}=f(X)$ and $\hat{Y}_k=P(Y_k=1|X)$, to minimize $\ell(f, X, Y) + \lambda R(f)$ where $\ell$ is the cross-entropy loss function and $R(f)$ is the regularizer of function $f$.

Minimizing the above objective function is a stochastic optimization problem. To solve it, we start with a batch setting to get an initialization of our machine learning model, and then utilize incremental learning techniques to refine our model in the online setting and adapt it to different operational modes of the machine. For the batch setting, we will first use coarse training data that could be properly labeled by machine and then obtain a pilot estimator by training a model on that sample. After the model is deployed, we will conduct incremental learning, which is similar to model fine tuning with \acrlong{SGD} (\acrshort{SGD}) method.

The major novelty of the proposed method is its real-time performance guarantee. In other words, solving the above optimization problem subject to a time constraint $t\leq T$. Since the algorithm will be implemented on an \acrshort{FPGA}, to accelerate model inference in a real-time, we propose two approaches for optimizing the \acrshort{ML} model in a resource-aware fashion: (i) We conduct model compression \cite{bucilua2006model} for existing models
%. Model compression is targeted to compress the machine learning model 
to lower their usage of computation resources 
%such as inference time 
and, at the same time, maintain original performance. For example, it can be realized by pruning inactive branches of neural networks to simplify the model \cite{han2015learning}, or by lowering precision of the model parameters \cite{gong2014compressing}. (ii) Furthermore, we will use \acrlong{NAS} (\acrshort{NAS}) \cite{zoph2016neural} methods to search for models with high enough performance and short enough inference time. \acrshort{NAS} is a meta-learning technique for automatically searching the optimal neural network. We could search for a model satisfying the time budget by limiting the number of \acrlong{FLOP}s (\acrshort{FLOP}s) in the objective function of \acrshort{NAS}.

\subsection{Model implementation and system architecture}
\label{sec:mlsystem}

To implement the powerful \acrshort{ML} algorithms developed in Sec.~\ref{sec:mlmodels}, we rely on {\bf co-design} -- the idea that system constraints, algorithm development, and hardware implementation inform and guide each other in complementary ways.
%Development of physics experiments are typically driven by this principle which starts from the physics requirements and drills down to the technical specifications of the apparatus. 
%Our co-design ecosystem is illustrated in Fig.~\ref{fig:codesign}. 
For accelerator operations, there are hard real-time latency constraints for very low latency processes. Therefore, we will explore hardware co-design for high-speed embedded technologies using \acrshort{FPGA}s. This project will develop a co-design methodology that is focused on providing cost-effective and highly tuned AI control systems with a quick turn-around time. 
\textbf{Co-PI Tran} and \textbf{Co-PI Memik} have strong expertise in developing optimized embedded FPGA systems for \acrshort{ML} algorithm implementations and other real-time applications.

\subsubsection*{Algorithm-Architecture Co-Design} \label{sec:HwAlgo}

% \begin{figure}[tbh]
% \centering
%   \includegraphics[keepaspectratio, width=0.6\textwidth]{LAB_20-2261_Embedded/images/codesign.png}
%   \caption{Hardware co-design principle which coherently integrates all levels of experimental design.}
%   \label{fig:codesign}
% \end{figure}

%The development of efficient \acrshort{AI} implementations in hardware begins with algorithm performance and design.
%As we consider how to implement the \acrshort{AI} hardware in high-rate and resource-constrained environments, 
%there are important considerations in building the most efficient systems. We must consider the size and structure of the \acrshort{AI} computation. 
This project will explore a variety of \acrshort{AI} methods (e.g., hierarchical and \acrshort{QoS}-driven reinforcement learning and supervised learning) to cater to the unique natures of the spill regulation and de-blending applications. However, we note that the co-design methodologies developed in this project will serve as templates for a large class of \acrshort{AI} methods in control design for experimental sciences. The common denominator is the underlying \acrshort{FPGA} \acrshort{SoC} hardware. The role of the co-design task is to construct a ubiquitous tool chain for mapping a variety of deep learning networks and their support systems (e.g., online learning module, communication interfaces, etc.) to the \acrshort{FPGA} \acrshort{SoC}, strategically re-organize resource allocation with an awareness of the target hardware platform's capabilities, and direct the design tools towards optimal system settings. 
%to deploy implementations of different neural networks resulting from these different \acrshort{AI} paradigms. 
The core computational module performing inference will be housed on the re-configurable logic of the \acrshort{FPGA} \acrshort{SoC}. Optimizing the performance of this module directly impacts the real-time performance goals of the system. Neural networks are generally characterized through a number of multiplication and addition operations using fitted parameters (weights) determined during the training procedure. By reducing both the number of mathematical operations and how often the weights need to be accessed, the implementation can be made more efficient. Further, the precision at which the calculations are performed is also important. 
%We explored the design space of \acrshort{AI} algorithm performance while reducing the calculation precision and removing unimportant calculations (pruning/compression). However, as neural network architectures become more complex and varied, this optimization must be understood in greater detail. New techniques, such as energy-aware pruning, can be used to improve the energy consumption directly~\cite{DBLP:journals/corr/YangCS16a}.
Just as important is to learn the most important features of the data for our challenge; learning the right representation as efficiently as possible can reduce computational complexity. 

{\bf Co-PI Memik} has extensive experience in developing analysis methods to identify performance bottlenecks in reconfigurable computing applications~\cite{8056827} and for machine learning applications in general~\cite{7363775}. As part of this task, we will perform analysis of the sensitivity of each learning model to the availability/scarcity or performance of a specific resource in our target device. For instance, for \acrshort{FPGA} hardware, width, depth, and connectivity of a network, precision of weights (resulting total storage and interconnect) will have varying correlations with the given capacity of different types of device resources (interconnect, embedded \acrshort{RAM}, embedded \acrshort{DSP} blocks used for multiply-accumulate operations, etc.). 

\subsubsection*{Programming Paradigms and Tools} \label{sec:Programming}

\begin{figure}[tbh]
\centering
  %\fbox
  {\includegraphics[keepaspectratio, width=0.8\textwidth]{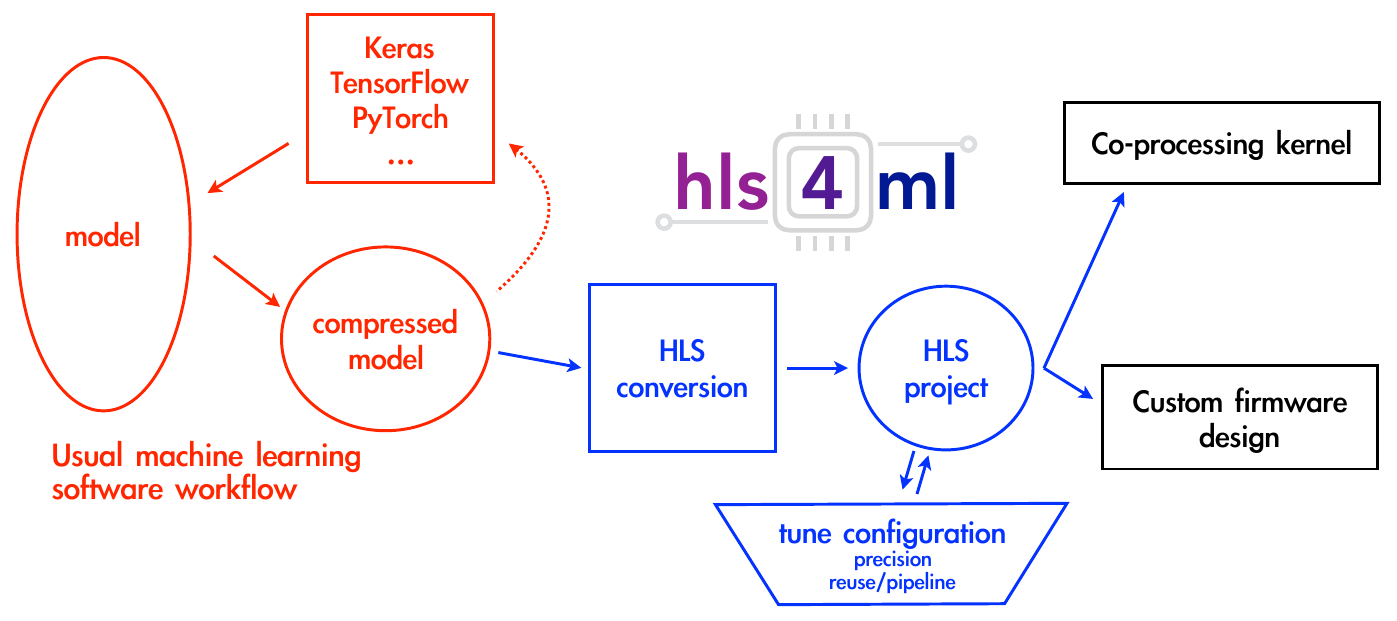}}
  \caption{Full example workflow from \acrshort{HLS} programming paradigms to hybrid solutions.}
  \label{fig:programming}
  \vskip-10pt 
\end{figure}
The main goal in developing programming paradigms and tools will be to increase accessibility of hardware implementations of  algorithms in order to accelerate the development cycles for \acrshort{AI} instrumentation. Mature programming tools are absolutely vital to wider deployment and adoption which will in turn improve the overall physics design process. 

We will align our tool development strategies with the unique aspects of \acrshort{AI} computation. One of the key features of neural networks is their modularity. This allows us to develop programming paradigms that enable the developer to separate and recombine these specific modular units to build larger neural network architectures. The basic description of the \acrshort{AI} circuit implementation, for example, can be described in a low level hardware description, but each kernel would be configurable based on resource, latency, and bandwidth constraints.   
 %Programming languages such as High Level Synthesis (HLS) and OpenCL are {\tt C}-based languages which are used to convert code into Register Transfer Level (RTL) Hardware Description Language (HDL). 

%%SEDA: I have written a modified version to clarify the confusion pointed out by reviewer BH. I have kept the original version in case you disagree. 

We will leverage a tool called {\tt hls4ml}~\cite{Duarte:2018ite, Summers:2020xiy, DiGuglielmo:2020eqx}, which takes popular open-source machine learning software frameworks such as {\tt TensorFlow}, {\tt Keras}, and {\tt PyTorch} and converts their model descriptions into \acrlong{HLS} (\acrshort{HLS}) code; e.g., C++-based code. \textbf{Co-PI Tran} is the leading developer of this framework.
The \acrshort{HLS} code is then converted by the Intel \acrshort{HLS} Compiler into a digital circuit implementation targeting the Arria10 FPGA. Unlike a pure C++ description of computation, the \acrshort{HLS} code is enhanced by special parameters (pragmas) which explicitly instruct the Intel \acrshort{HLS} Compiler to tune the performance of the underlying hardware description to customize it for different system constraints. The full workflow of {\tt hls4fml} is depicted in Fig.~\ref{fig:programming}. Despite being a new software package, this tool has seen widespread adoption in the \acrshort{HEP} community and has been successfully used for fully-connected neural networks in \acrshort{LHC} trigger applications on \acrshort{FPGA}s. {\bf Co-PI Memik} will leverage her expertise on design automation tools for reconfigurable architectures~\cite{968653,903440, 4427446} and extend the capabilities of the Intel \acrshort{HLS} Compiler through systematic exploration of the \acrshort{HLS} pragmas.

\subsubsection*{System Design and Integration} \label{sec:Integration}
Solutions will be needed for integration of the \acrshort{AI} implementations into a coherent accelerator control system, including: {\bf interfacing} \acrshort{AI} kernels to components such as power and memory management, device infrastructure and controls, and networking protocols; {\bf communication} with other devices in both a homogeneous and heterogeneous hardware stack; {\bf software interfaces} with the operators and users for features such as neural network weight programmability and neural network training and feedback loops. {\bf Co-PI Memik} has past experience on design of real-time applications on \acrshort{FPGA}s~\cite{1515761,5694230} and co-designing reconfigurable  hardware/software systems~\cite{4397281}. She will explore the optimal partitioning of the integration components within the Intel Arria10 \acrshort{SoC}.

An example of a conceptual system, which includes the three aspects listed above, is shown in Fig.~\ref{fig:rl}.
Illustrated here is a dynamic reinforcement learning architecture for a generic system which controls an experimental apparatus. This topology serves as a realistic representation for the applications considered in this project. The main component in this system is the Intel Arria10 \acrshort{SoC}. This system communicates with a high-speed data acquisition system which aggregates data in long term storage for offline model training as well as for the development of a complex digital twin. As the coarse grain features of the models are discovered through offline training on \acrshort{GPU}s, incremental updates will be performed on the model through the dynamic feedback loop within the Intel Arria10 \acrshort{SoC}.  

The target model for controlling the experiment is shown in the red box labeled ``Target Model''. Batches of data are recorded and used to continue training the model based on a desired reward. The target model will require an interface which provides a mechanism to update the weights via the ARM system responsible for training on real-time data. Updating the target model in this manner will require a thoughtful approach to system integration and \acrshort{FPGA} firmware infrastructure. While this is but a simplified example, it serves to illustrate the interfaces required to implement a configurable embedded \acrshort{ML} system.

%Finally, and in addition to the interfaces described above, firmware will be required for the execution of inference. For instance, the \acrshort{ML} model will be interfaced with a data queue, which organizes sensor readings from the \acrshort{BLM}s. At this interface, a structure to correctly map each \acrshort{BLM} to the corresponding node in the input layer will need to be established.  

% \begin{figure}[tbh]
% \centering
%   \includegraphics[keepaspectratio, width=0.8\textwidth]{LAB_20-2261_Embedded/images/rl-workflow.png}
%   \caption{An example of a dynamically re-configurable system}
%   \label{fig:reinforce}
% \vskip-10pt
% \end{figure} 

\section{Timetable of Activities}

% {\bf Timeline for all major activities including milestones and deliverables.}\\
\begin{figure}[tbh]
\centering
  \includegraphics[keepaspectratio, width=0.9\textwidth]{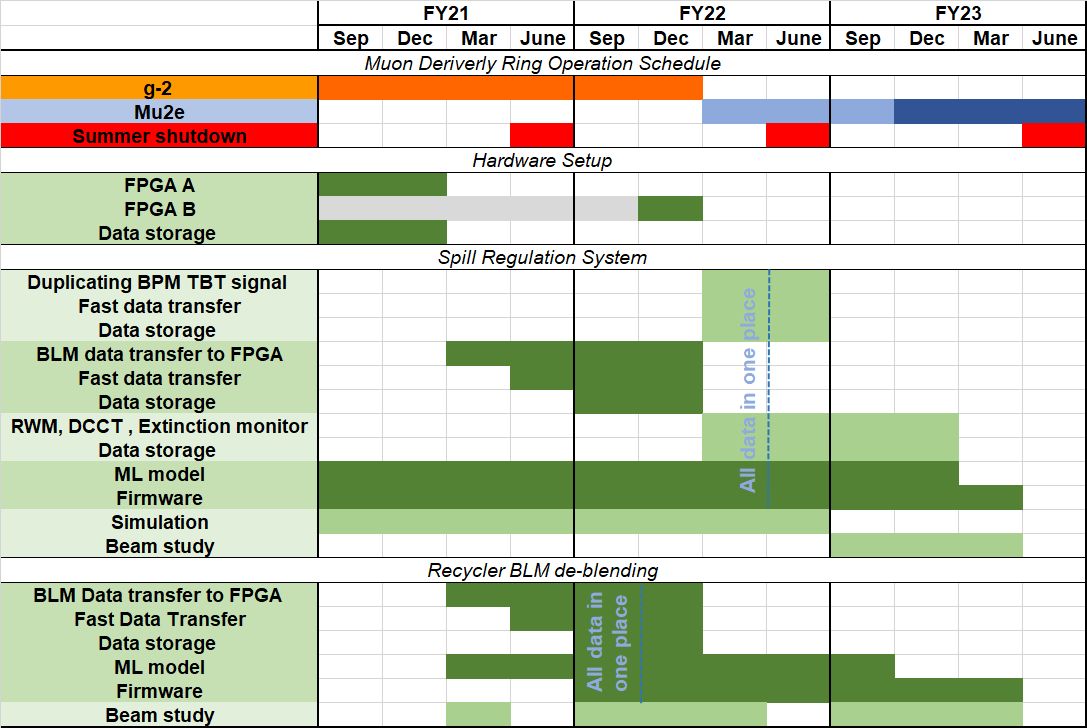}
  \caption{Project timeline and milestones.}
  \label{fig:timeline}
\end{figure}

% Discuss risks and mitigation strategies.

This proposal requests three years of support to complete two main projects: integrating \acrshort{ML} into the spill regulation system and de-blending the \acrshort{MI} and \acrshort{RR} \acrshort{BLM} signals.  Table 1 shows the timeline for our proposal and includes the accelerator operation plan.  This schedule assumes that funding begins in September 2020 and that the accelerator complex will continue to have a 12-week summer shutdown for the next three years. The ongoing experiment at the Muon Campus, g-2, is predicted to end by March 2022.  Beam commissioning for the \acrshort{Mu2e} Slow Extraction will start in the Delivery Ring in late 2022.  At this time low intensity beams will be slowly extracted to the temporary dump on a regular basis which can be used for beam studies of the \acrshort{SRS}. Dedicated time for \acrshort{ML} studies will easily fit into the schedule of machine development. During the operation of the g-2 experiment, it is expected that the \acrshort{RR} will deliver beam with intensities and injection patterns similar to both g-2 and \acrshort{Mu2e} experiments, enabling the development of \acrshort{Mu2e} \acrshort{ML} beam studies. 

The work packages in this proposal fall into 5 different types: (i) extract low latency signals from instrumentation; (ii) transmit and store the data elements; (iii) conduct beam studies and simulation to 
%train the ML model; 
collect training data for the \acrshort{ML} models; (iv) build and train ML models, surrogate models and an online agent; (v) implement the \acrshort{ML} model on the \acrshort{FPGA} and set up an online training system, coordinating the \acrshort{FPGA} and the embedded CPU of the \acrshort{SoC} package. The project work packages and their timeline is shown in Fig.~\ref{fig:timeline}. Further, project milestones based on these work packages are shown in Table \ref{table:tmln}.

\begin{table}[tbh]
%\small
\begin{tabular}{|p{2cm}||p{13cm}|}
\hline
                  \multicolumn{2}{|l|}{\textbf{Spill Regulation System}} \\ \hline \cline{2-2} 
                  \multirow{2}{*}{FY21} &  Transmit existing BLM system data to the FPGA board                          \\ \cline{2-2} 
 & Build ML models using an analytical model of the spill regulation loop.                          \\ \hline
  \multirow{2}{*}{FY22}                & Transmit BPM/BLM system data to centralized FPGA board, then transmit to data storage.                          \\ \cline{2-2} 
& Establish ML models, surrogate models and online agent, for controls using all input signals and data in the storage. \\ \hline
                  & Implement the ML model into the FPGA board                         \\ \cline{2-2} 
                  & Test the spill regulation loop with ML.                         \\ \cline{2-2} 
\multirow{-4}{*}{FY23} & Conduct beam studies and measurements, study ML performance.                          \\ \hline
\multicolumn{2}{|l|}{\textbf{MI/RR BLM de-blending}} \\ \hline \cline{2-2} 
                  &  Transmit existing BLM system data to the FPGA board from 7 nodes                         \\ \cline{2-2} 
\multirow{-2}{*}{FY21} & Conducting beam studies and measurements and accumulating data for training.                         \\ \hline
                  &   Establishing ML models, surrogate models and online agent, for controls using all input signals and data in the storage.                       \\ \cline{2-2} 
\multirow{-4}{*}{FY22} & Implementing the ML model into the centralized board, FPGA board                         \\ \hline
                  &  Conduct beam studies, and compare the results between original system and the one with ML.                         \\ \cline{2-2} 
\multirow{-4}{*}{FY23} &   Set up necessary control parameters and monitors and replacing the machine permit consolidation system with the new ML system for operation.                       \\ \hline
\end{tabular}
\caption{Project Milestones}
\label{table:tmln}
\normalsize
\end{table}

\section{Project Management Plan}

%{\bf Multi-institutional proposals must include a project management plan that clearly indicates the roles and responsibilities of each organization and indicates how activities will be coordinated and communicated among team members.}\\

%Put together a little management team together.

\begin{figure}[tbh]
\centering
%\vskip-15pt 
  %\fbox
  {\includegraphics[keepaspectratio, width=0.7\textwidth]{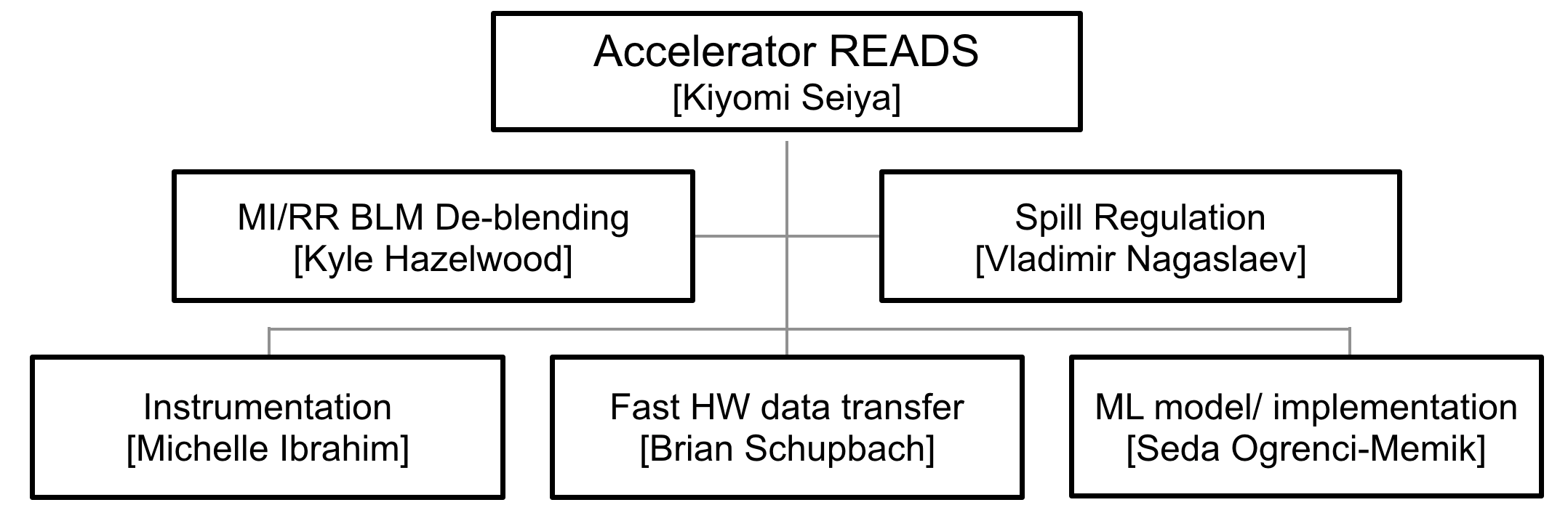}}
  \caption{Proposal management structure}
  \label{fig:orgchart}
  %\vskip-10pt 
\end{figure}

This proposal is composed of two primary deliverables: MI/RR BLM de-blending and Slow spill regulation. 
It brings together a strong multi-disciplinary team with accelerator physicists, beam
instrumentation engineers, embedded system architects, \acrshort{FPGA} board design experts and \acrlong{ML} experts.
The project management plan is presented in Fig.~\ref{fig:orgchart} and has two physics focus areas and three technical focus areas with coordinators for each.   
The multidisciplinary project team will consist of Fermilab staff crossing division boundaries and collaborators from Northwestern University. 
Furthermore, the proposal provisions for effort from instrumentation engineers, controls and computer engineers, an ML research associate, and graduate students to drive the work plan under the management team.
They will receive the support and mentoring by other staff members to ensure this project will be an informative experience that will advance them along their career paths.
Experienced and knowledgeable experts lead the physics and technical teams and ensure our goal is achieved in three years.  
% Engineer A and the graduate students who focus on this project for three years will receive the support and mentoring by other staff members to ensure this project will be an informative experience that will advance them along their career paths.  

The roles and responsibilities of the individuals are as follows:\\
\noindent {\bf Lead PI: Kiyomi Seiya}, Senior Scientist in the Accelerator Division (AD), who has 20 years’ experience in Fermilab AD operation and beam physics, organizes the team, coordinates the schedule and budget and ensures that this project aligns with the laboratory’s schedule. 

\noindent {\bf Accelerator Physicists: Kyle Hazelwood}, Engineering Physicist III in the AD Main Injector department, has been working on AD operations, accelerator controls, and beam physics for 13 years and leads MI/RR BLM De-blending. He is responsible for beam studies and analysis and coordinates the required hardware and software with the technical development teams. He will collaborate with the technical development teams on \acrshort{ML} model development.  
{\bf Vladimir Nagaslaev}, Senior Scientist in AD for 18 years, who has been responsible for design and implementation of the Slow Extraction for the Mu2e project, leads Spill Regulation. 

\noindent {\bf Instrumentation: Michelle Ibrahim}, Senior engineer in the AD Instrumentation department, leads the Instrumentation team. Ibrahim has been developing the spill regulation system for the Mu2e project and is an expert on signal processing for the beam diagnostics system.  Engineer A, who is a new hire, focuses on code development on the FPGA board which receives data from the existing system under Ibrahim’s supervision.  Engineer A also digitizes, and processes all input signals into an FPGA board which will be used for the spill regulation system.  

\noindent{\bf Hardware and Fast Data Transfer: Brian Schupbach}, Staff Engineer and an Low Level RF expert, leads the Hardware and Fast Data Transfer team.  Schupbach developed an FPGA board for a Fermilab LDRD project to implement ML into a bending magnet power supply regulation system.  Schupbach is responsible for all FPGA board development and necessary modification. {\bf Dennis Nicklaus} Software Engineer, will develop fast communication protocols via ethernet which allow for fast data transfer between multiple FPGA boards with Schupbach.  Nicklaus will be responsible for testing fast communication between the distributed systems. 

\noindent {\bf ML Model and Firmware: Seda Ogrenci-Memik}, Professor at Northwestern University, leads the ML model and Firmware team and implements ML models into fast embedded systems which are based on the Arria10 \acrshort{FPGA} \acrshort{SoC} system with graduate student A. She will collaborate with Co-PI Tran on the embedded systems design and automated tool development to support the design efforts of the project. She will also collaborate with Co-PI Liu on  hardware/resource-aware \acrshort{ML} model compression. 
{\bf Nhan Tran}, Scientist in AI, is an expert on fast embedded system support hardware and software development and tracks the progress of the project.  He will support a Fermilab AI research associate.  He will collaborate with Co-PI Memik on ML hardware implementations and Co-PI Liu on ML model development.  
{\bf Han Liu}, professor at Northwestern University, develops ML Models, surrogate models and an online agent with graduate student B. A research associate in Al works on both ML modeling and firmware development with the collaborators.

\section{Project Objectives}

The overarching goal of our proposal, Accelerator Real-time Edge AI Distributed System (Accelerator READS) is to combine a 'Fast embedded ML' system and a 'Fast data transfer system' to improve the operation of large accelerator complexes by means of global ML based supervision.  We will demonstrate these techniques with two experiments of significance: \emph{the Mu2e spill regulation system} and \emph{the de-blending of MI and RR beam losses}.

In the first instance we aim to increase the Spill Duty Factor of slow spill extraction with the use of both model-based and model-free reinforcement learning algorithms to assist the spill regulation in real-time increase the uptime. In the second instance, supervised learning algorithms will be used to  decouple overlapping beam losses in the MI/RR enclosure to reduce beam aborts.

The Mu2e construction phase is currently nearing completion; Mu2e is expected to come online as the g-2 experiment winds down.  On a relatively short time scale, Machine Learning techniques have the  potential to significantly enhance the experimental physics output by improving the performance of beam delivery.
 
Successfully implementing ML in a fast-embedded systems environment will require the development of a data processing pipeline, hardware and software. The process is as follows: 
(i) development of control modules based on the Arria10 SOC FPGA system to allow for the reception of data sets from the existing instrumentation and for the extraction of low latency signals to be used as training inputs for ML. 
(ii) development of a high-speed data transmission protocol between distributed FPGA modules and of associated infrastructure to transmit data elements to central modules and put them into data storage; this includes a computing server for user interface, data storage, reprocessing, and  manipulation.
(iii) conduct beam studies and simulation to support ML model development, build ML models, surrogate models and an online agent.
(iv) implement the ML model on central modules and set up an online training system by developing code using hls4ml.

A collaboration between Fermilab and Northwestern University will synthesize the talents and resources of accelerator physicists, beam instrumentation engineers, embedded system architects and ML experts to meet the challenges related to complex real-time accelerator controls. A new hired engineer and two graduate students from NU will focus on the project for three years and will receive support from staff members. This environment will provide a singular learning experience that will advance them along their career paths. The timeline was carefully examined to ensure that  the proposed schedule is well-aligned with both the AD operation and Mu2e experiment schedules. Experienced and knowledgeable experts will lead the physics and technical teams to ensure our goal is achieved in  three years. 

\textbf{The techniques described in this proposal will bring a new and unique capability to the accelerator facility and develop methods which can improve operations at other large accelerator complexes and future high intensity operation. 
}

\clearpage

\newpage
\appendix

% Use arabic sectioning in the appendix
\renewcommand\thesection{\arabic{section}}
\renewcommand\thesubsection{\thesection.\arabic{subsection}}
\fancyfoot[LO,LE]{Appendix \thesection}

% Appendix 1
%\section{BIOGRAPHICAL SKETCHES}
%\input{bios}
%\newpage

% Appendix 2
%\section{CURRENT AND PENDING SUPPORT}
%\input{support}
%\newpage

% Appendix 3
\section{BIBLIOGRAPHY \& REFERENCES CITED}
% APPENDIX 3: BIBLIOGRAPHY & REFERENCES CITED Provide a bibliography

%\newpage
%\bibliographystyle{abbrv}
%\bibliographystyle{unsrt}
\bibliographystyle{IEEEtran}
\bibliography{bib/references}

\begin{thebibliography}{10}
\providecommand{\url}[1]{#1}
\csname url@samestyle\endcsname
\providecommand{\newblock}{\relax}
\providecommand{\bibinfo}[2]{#2}
\providecommand{\BIBentrySTDinterwordspacing}{\spaceskip=0pt\relax}
\providecommand{\BIBentryALTinterwordstretchfactor}{4}
\providecommand{\BIBentryALTinterwordspacing}{\spaceskip=\fontdimen2\font plus
\BIBentryALTinterwordstretchfactor\fontdimen3\font minus
  \fontdimen4\font\relax}
\providecommand{\BIBforeignlanguage}[2]{{%
\expandafter\ifx\csname l@#1\endcsname\relax
\typeout{** WARNING: IEEEtran.bst: No hyphenation pattern has been}%
\typeout{** loaded for the language `#1'. Using the pattern for}%
\typeout{** the default language instead.}%
\else
\language=\csname l@#1\endcsname
\fi
#2}}
\providecommand{\BIBdecl}{\relax}
\BIBdecl

\bibitem{ml}
A.~Edelen \emph{et~al.}, ``Opportunities in machine learning for particle
  accelerators,'' \emph{arXiv preprint arXiv:1811.03172}, 2018.

\bibitem{Duarte:2018ite}
J.~Duarte \emph{et~al.}, ``{Fast inference of deep neural networks in FPGAs for
  particle physics},'' \emph{JINST}, vol.~13, no.~07, p. P07027, 2018.

\bibitem{Bartoszek:2014mya}
\BIBentryALTinterwordspacing
L.~Bartoszek \emph{et~al.}, ``{Mu2e Technical Design Report},'' 2014. [Online].
  Available: \url{https://arxiv.org/abs/1501.05241}
\BIBentrySTDinterwordspacing

\bibitem{sdf_c}
\BIBentryALTinterwordspacing
{V. Kain \it et al.,}. (2019) “\uppercase{R}esonant slow extraction with
  constant optics for improved separatrix control at the extraction septum".
  [Online]. Available:
  \url{https://doi.org/10.1103/PhysRevAccelBeams.22.101001}
\BIBentrySTDinterwordspacing

\bibitem{sdf_j}
\BIBentryALTinterwordspacing
{D. Naito \it et al.} (2019) “\uppercase{R}eal-time correction of betatron
  tune ripples on a slowly extracted beam. [Online]. Available:
  \url{https://doi.org/10.1103/PhysRevAccelBeams.22.072802}
\BIBentrySTDinterwordspacing

\bibitem{Ainsworth:2018ntr}
R.~Ainsworth, P.~Adamson, B.~Brown, D.~Capista, K.~Hazelwood, I.~Kourbanis,
  D.~Morris, M.~Xiao, and M.-J. Yang, ``{High Intensity Proton Stacking at
  Fermilab: 700 kW Running},'' in \emph{{61st ICFA Advanced Beam Dynamics
  Workshop on High-Intensity and High-Brightness Hadron Beams}}, 2018, p.
  TUA1WD04.

\bibitem{dr_re}
\BIBentryALTinterwordspacing
{V. Nagaslaev \it et al.,}. (2011) “\uppercase{T}hird integer resonance slow
  extraction using \uppercase{RFKO} at high space charge". [Online]. Available:
  \url{https://www.osti.gov/biblio/1031169}
\BIBentrySTDinterwordspacing

\bibitem{bertsekas1996neuro}
D.~P. Bertsekas and J.~N. Tsitsiklis, \emph{Neuro-dynamic programming}.\hskip
  1em plus 0.5em minus 0.4em\relax Athena Scientific, 1996.

\bibitem{barto2003recent}
A.~G. Barto and S.~Mahadevan, ``Recent advances in hierarchical reinforcement
  learning,'' \emph{Discrete event dynamic systems}, vol.~13, no. 1-2, pp.
  41--77, 2003.

\bibitem{bucilua2006model}
C.~Buciluǎ, R.~Caruana, and A.~Niculescu-Mizil, ``Model compression,'' in
  \emph{Proceedings of the 12th ACM SIGKDD international conference on
  Knowledge discovery and data mining}, 2006, pp. 535--541.

\bibitem{han2015learning}
S.~Han, J.~Pool, J.~Tran, and W.~Dally, ``Learning both weights and connections
  for efficient neural network,'' in \emph{Advances in neural information
  processing systems}, 2015, pp. 1135--1143.

\bibitem{gong2014compressing}
Y.~Gong, L.~Liu, M.~Yang, and L.~Bourdev, ``Compressing deep convolutional
  networks using vector quantization,'' \emph{arXiv preprint arXiv:1412.6115},
  2014.

\bibitem{zoph2016neural}
B.~Zoph and Q.~V. Le, ``Neural architecture search with reinforcement
  learning,'' \emph{arXiv preprint arXiv:1611.01578}, 2016.

\bibitem{8056827}
Y.~{Luo}, X.~{Wen}, K.~{Yoshii}, S.~{Ogrenci-Memik}, G.~{Memik}, H.~{Finkel},
  and F.~{Cappello}, ``Evaluating irregular memory access on opencl fpga
  platforms: A case study with xsbench,'' in \emph{2017 27th International
  Conference on Field Programmable Logic and Applications (FPL)}, 2017, pp.
  1--4.

\bibitem{7363775}
S.~M. {Faisal}, G.~{Tziantzioulis}, A.~M. {Gok}, N.~{Hardavellas},
  S.~{Ogrenci-Memik}, and S.~{Parthasarathy}, ``Edge importance identification
  for energy efficient graph processing,'' in \emph{2015 IEEE International
  Conference on Big Data (Big Data)}, 2015, pp. 347--354.

\bibitem{Summers:2020xiy}
\BIBentryALTinterwordspacing
S.~Summers \emph{et~al.}, ``{Fast inference of Boosted Decision Trees in FPGAs
  for particle physics},'' 2 2020. [Online]. Available:
  \url{https://arxiv.org/abs/2002.02534}
\BIBentrySTDinterwordspacing

\bibitem{DiGuglielmo:2020eqx}
\BIBentryALTinterwordspacing
V.~Loncar \emph{et~al.}, ``{Compressing deep neural networks on FPGAs to binary
  and ternary precision with HLS4ML},'' 3 2020. [Online]. Available:
  \url{https://arxiv.org/abs/2003.06308}
\BIBentrySTDinterwordspacing

\bibitem{968653}
S.~{Ogrenci Memik}, E.~{Bozorgzadeh}, R.~{Kastner}, and M.~{Sarrafzadeh}, ``A
  super-scheduler for embedded reconfigurable systems,'' in \emph{IEEE/ACM
  International Conference on Computer Aided Design. ICCAD 2001. IEEE/ACM
  Digest of Technical Papers (Cat. No.01CH37281)}, 2001, pp. 391--394.

\bibitem{903440}
K.~{Bazargan}, R.~{Kastner}, S.~{Ogrenci}, and M.~{Sarrafzadeh}, ``A c to
  hardware/software compiler,'' in \emph{Proceedings 2000 IEEE Symposium on
  Field-Programmable Custom Computing Machines (Cat. No.PR00871)}, 2000, pp.
  331--332.

\bibitem{4427446}
M.~{Santambrogio}, M.~{Giani}, and S.~O. {Memik}, ``Managing reconfigurable
  resources in heterogeneous cores using portable pre-synthesized templates,''
  in \emph{2007 International Symposium on System-on-Chip}, 2007, pp. 1--4.

\bibitem{1515761}
D.~{Nguyen}, G.~{Memik}, S.~O. {Memik}, and A.~{Choudhary}, ``Real-time feature
  extraction for high speed networks,'' in \emph{International Conference on
  Field Programmable Logic and Applications, 2005.}, 2005, pp. 438--443.

\bibitem{5694230}
B.~{Leung}, C.~{Wu}, S.~O. {Memik}, and S.~{Mehrotra}, ``An interior point
  optimization solver for real time inter-frame collision detection: Exploring
  resource-accuracy-platform tradeoffs,'' in \emph{2010 International
  Conference on Field Programmable Logic and Applications}, 2010, pp. 113--118.

\bibitem{4397281}
M.~D. {Santambrogio}, V.~{Rana}, S.~O. {Memik}, U.~A. {Acar}, and D.~{Sciuto},
  ``A novel soc design methodology combining adaptive software and
  reconfigurable hardware,'' in \emph{2007 IEEE/ACM International Conference on
  Computer-Aided Design}, 2007, pp. 303--308.

\end{thebibliography}
%\bibliography{bib/bibentries,bib/hepcloud,bib/obama,bib/process,bib/projectq,bib/ornl}

% render the glossary
\printnoidxglossaries

\newpage

% Appendix 4
\section{FACILITIES}

%{\sc\large\bf FACILITIES, EQUIPMENT, AND OTHER RESOURCES} \\
% \subsection{FERMI NATIONAL ACCELERATOR LABORATORY}

%This information is used to assess the capability of the organizational resources, including subawardee resources, available to perform the effort proposed. Identify the facilities to be used (Laboratory, Animal, Computer, Office, Clinical and Other). If appropriate, indicate their capacities, pertinent capabilities, relative proximity, and extent of availability to the project. Describe only those resources that are directly applicable to the proposed work. Describe other resources available to the project (e.g., machine shop, electronic shop) and the extent to which they would be available to the project. For proposed investigations requiring access to experimental user facilities maintained by institutions other than the applicant, please provide a document from the facility manager confirming that the researchers will have access to the facility. Please provide the Facility and Other Resource information as an appendix to your project narrative.

\subsection*{Fermi National Accelerator Laboratory}

Fermilab operates a large accelerator facility consisting of 35~keV preacc, 400~MeV H- linac, 8~GeV Booster proton synchrotron, and 120~GeV Main Injector synchrotron, two 8~GeV storage rings, Recycler and Delivery Ring, and their beam lines.  These accelerators and their control systems will be used in this proposal.

Fermilab provides facilities computing resources in support of it's HEP mission, for archival data storage, high throughput (grid) computing and networking.  
The Fermilab high throughput grid computing facilities operates computing systems providing 19,664 x86 based computing cores with an annual capacity to provide approximately 172 million CPU hours.  The facility provides gigabit Ethernet connectivity within the grid clusters and provides higher aggregated bandwidth paths to central storage facilities.   The facility is shared across Fermilab hosted and associated experiments through a HTCondor based batch submission system.  This system permits the central data processing/analysis teams of each experiment to conduct large scale simulation and data analysis on their experiment's datasets.  
%The facility also provides the capabilities to allow individual experimenters associated with Fermilab experiments to submit their workloads and receive back their results.  This facility serves as the basis for current FNAL analysis work for intensity frontier experiments (NOvA, MicroBooNE, DUNE) and for the USCMS Tier 1 facility.

The computing capabilities of the Fermilab facility are augmented by the storage systems that the laboratory operates.  The storage facility features four, ten thousand (10,000) slot Oracle SL8500 robotic tape libraries with an additional 3 tape libraries dedicated to the CMS experiment.  The storage facility operates 69 tape drives supporting T10Kc, T10Kd, and LTO4 media.  The facility has on the order of 15,000 active media cartridges with an additional 16,000 slots occupied by data migration processes. The facility archival storage facility's tape systems are fronted by a distributed disk caching system with provides 3.4~PB of read/write cache buffer from/to tape, a 1.4~PB non-tape backed cache for large scale data analysis operations.  %This facility is the standard data archive and repository for data for the NOvA, MicroBooNE, DUNE and other intensity frontier experiments.  The facility is also the storage base of the CMS Tier 1 facility. 

\subsection*{Northwestern University: Department of Electrical and Computer Engineering and Department of Computer Science}

The Department of Electrical and Computer Engineering (ECE) and The Department of Computer Science (CS) are part of the Robert R. McCormick School of Engineering and Applied Science of Northwestern University. ECE consists of 27 full-time faculty members and 3 faculty members with joint appointments. CS consists of 34 full-time faculty members. Many faculty members are fellows of their professional societies. Faculty are fellows with IEEE, ACM, AAAS, AAAI, APS, APA, OSA, MRS, SPIE, AIMBE, The Cognitive Science Society, and the Human Factors and Ergonomics Society. The majority of the junior faculty members have received prestigious young investigator awards, such as the NSF CAREER award. Both ECE and CS are highly interdisciplinary department, with faculty members collaborating across Northwestern and with other institutions.

ECE Department laboratories and classrooms are located in the 750,000 square-feet Technological Institute, which houses the McCormick School of Engineering. CS Department laboratories and classrooms are located in the newly renovated Seeley Mudd Building providing 22,500 square feet dedicated to the growing Computer Science program. PIs’ students have offices in the Tech Institute or Mudd Hall respectively. The two buildings are connected by an internal bridge that is only a short walk in distance
\newpage

% Appendix 5
\section{EQUIPMENT}

% APPENDIX 5: EQUIPMENT

% List major items of equipment already available for this project
% and, if appropriate identify location and pertinent capabilities.
% Provide the Equipment information as an appendix to your project
% narrative.

% \subsection{FERMI NATIONAL ACCELERATOR LABORATORY}

\subsection*{Fermi National Accelerator Laboratory}
The Accelerator READS proposal will utilize the existing Accelerator Division controls infrastructure (existing front-end devices, physical network lines and communication nodes, etc.).
Our solutions will be parasitically integrated into existing beam instrumentation readout electronics hardware and their existing infrastructure will be used.  The development of \acrshort{SoC} embedded hardware systems will utilize electronics lab space and tools for testing and validation of their performance.

Scientific Computing Division infrastructure will be used in the central data storage setup and utilization.

\subsection*{Northwestern University: Department of Electrical and Computer Engineering and Department of Computer Science}

The departments have access to a large number of servers, as well as a parallel and distributed infrastructure. Servers range from 8-node to 128-node clusters with large memories and many terabytes of disk space. Furthermore, there are specialized clusters, such as FPGA clusters, and NVIDIA GP/GPU servers. In collaboration with the McCormick School of Engineering and NUIT, we have established two full featured machine rooms in the main engineering building. There are several racks of machines supporting various research projects in these spaces.
NUIT manages a large High-Performance Computing Cluster (HPCC) called Quest with, currently, over 11,800 CPU cores and at least 96GB of memory per core.

The departments also have access to tera- and peta- scale supercomputing resources through its research collaborations with Argonne National Laboratory and affiliation with the Great Lake Consortium for Petascale Computation (GLCPC), led by the National Center for Supercomputing Applications (NCSA). In particular, the departments have access to various processing clusters (several at NCSA) and IBM Blue Gene supercomputers. Several faculty also have access (via NSF and DOE allocations) to supercomputers at Oak Ridge National Laboratory, San Diego Supercomputer Center, Texas Advanced Computing Center (TACC), and Lawrence Berkeley National Laboratory.

\newpage

% Appendix 6
\section{DATA MANAGEMENT PLAN}

% APPENDIX: DATA MANAGEMENT PLAN
\noindent For the budget years in this proposal, this proposal will produce data from the following sources 
\begin{itemize}
    \item Raw data from testing and calibration of beam instrumentation front-end electronics 
    \item Simulated data generated for machine learning model fine-tuning
\end{itemize}
A data storage pipeline will be used to process the raw data to produce
physics measurements and persisted using ROOT-based and HDF5-based data models. 
All data will be centrally stored at FNAL and be made available to all members of the proposal.  

\subsubsection*{Plan for Serving Data to the Collaboration and Community}

Before being released to the collaboration, data is tagged using the code version used to produce it.  These tagged releases will serve as the standard data sets that will be used for analysis and publication.  Dissemination of the data beyond collaborators will be resource prohibitive.

\subsubsection*{Plan for Making Data Used in Publications Available}

In all cases of publications, data in the plots, charts, and figures, and 
Digital Object Identifiers will be made available in accordance with policy at
the time of publication by using mechanisms provided by the publisher, hosting 
by a collaborating institution or services provided by INSPIRE. This includes
publications resulting from research data from experiments, simulation, and 
research and development projects such as detector prototype data.

\subsubsection*{Responsiveness to Office of Science Statement on Digital Data Management}

The data management plan fully adheres to the recently implemented policy of the 
DOE Office Science: {http://sciences.energy.gov/funding-opportunities/digital-data-management},
except that not all data is planned to be publicly available due to resource
limitations.

\newpage

% Appendix 7
%\section{LETTERS OF COMMITMENT}
%\includepdf[pages=-,pagecommand={\thispagestyle{plain}}]{Partner_Commitment_Letter.pdf}
%\newpage

% Appendix 8 -- OPTIONAL
%\section{OTHER ATTACHMENTS}
%\includepdf[pages=-,pagecommand={\thispagestyle{plain}}]{Other_Attachment.pdf}
%\newpage

\end{document}